\documentclass[reprint,amsmath,amssymb,aps,pre,article]{revtex4-2}
\usepackage{graphicx}
\usepackage{dcolumn}
\usepackage{bm}
\usepackage{xcolor}
\usepackage{float}
\usepackage{mathrsfs}
\usepackage{lipsum}
\usepackage{dsfont}
\usepackage{hyperref}
\usepackage{blindtext}

\begin{document}

\preprint{APS/123-QED}
\title{Dynamics of reservoir computing for crises prediction}

\author{Dishant Sisodia}\email{dishantsisodia.physics@gmail.com}
\author{Sarika Jalan} \email{sarika@iiti.ac.in: Corresponding author}
\affiliation{Complex Systems Lab, Department of Physics, Indian Institute of Technology Indore, Khandwa Road, Simrol, Indore-453552, India}

\date{\today}

\begin{abstract}
Reservoir computing has emerged as a powerful framework for time series modeling and forecasting including the prediction of the discontinuous transitions. However, the mechanism behind its success is not yet fully understood. This Letter elucidates the functioning of reservoir computing by examining its successful prediction of boundary and attractor-merging crises. 
We investigate in detail how the reservoir’s internal dynamics mimic the actual system, that enables it to accurately reproduce the scaling exponent near boundary crisis. We establish this across distinct systems, exemplified by the logistic and Gauss maps.
The study contributes to a broader understanding of the internal dynamics that enable learning algorithms to anticipate critical transitions.
\end{abstract}

\maketitle

\paragraph{\bf{Introduction:} }
Machine learning has become an integral part of complex systems research \cite{Tsironis2025} with diverse applications such as discovering model equations from data \cite{Brunton2016}, predicting discrete bifurcations using deep learning \cite{Bury2023}, Hamiltonian neural networks \cite{Choudhary2020}, fatigue detection using EEG datasets \cite{10.1063/1.5120538}, etc. Understanding the mechanisms by which machines make accurate predictions from a dynamical perspective is an emerging research area that guides the development of more robust and better algorithms. Reservoir computing (RC) is a machine learning approach built upon the framework of recurrent neural networks (RNNs) that has recently drawn good attention 
\cite{LUKOSEVICIUS2009127,doi:10.1126/science.1091277}, particularly, due to its suitability for handling sequential tasks, such as forecasting chaotic time series \cite{10.1063/1.5039508}. Further, if a systems parameter is included as an additional input, RC has been able to reproduce entire dynamics of the system in the neighborhood of its transition points \cite{PhysRevResearch.3.013090,PhysRevE.110.034211}. This capability is especially valuable since governing equations are rarely known in practice. 

Despite the remarkable predictive success of RC, the mechanism behind its performance remains poorly understood with only a few recent studies confined to a purely mathematical viewpoint, lacking motivation from nonlinear-dynamics perspectives. \cite{Hart2020,Bollt2021}.  Earlier studies employed bifurcation analysis by tracking eigenvalue motion to account for the accurate prediction of transition to amplitude death
\cite{PhysRevE.110.034211}.

\begin{figure}[t!]
\includegraphics[scale=0.28, trim= 9cm 3cm 0 2cm, clip]{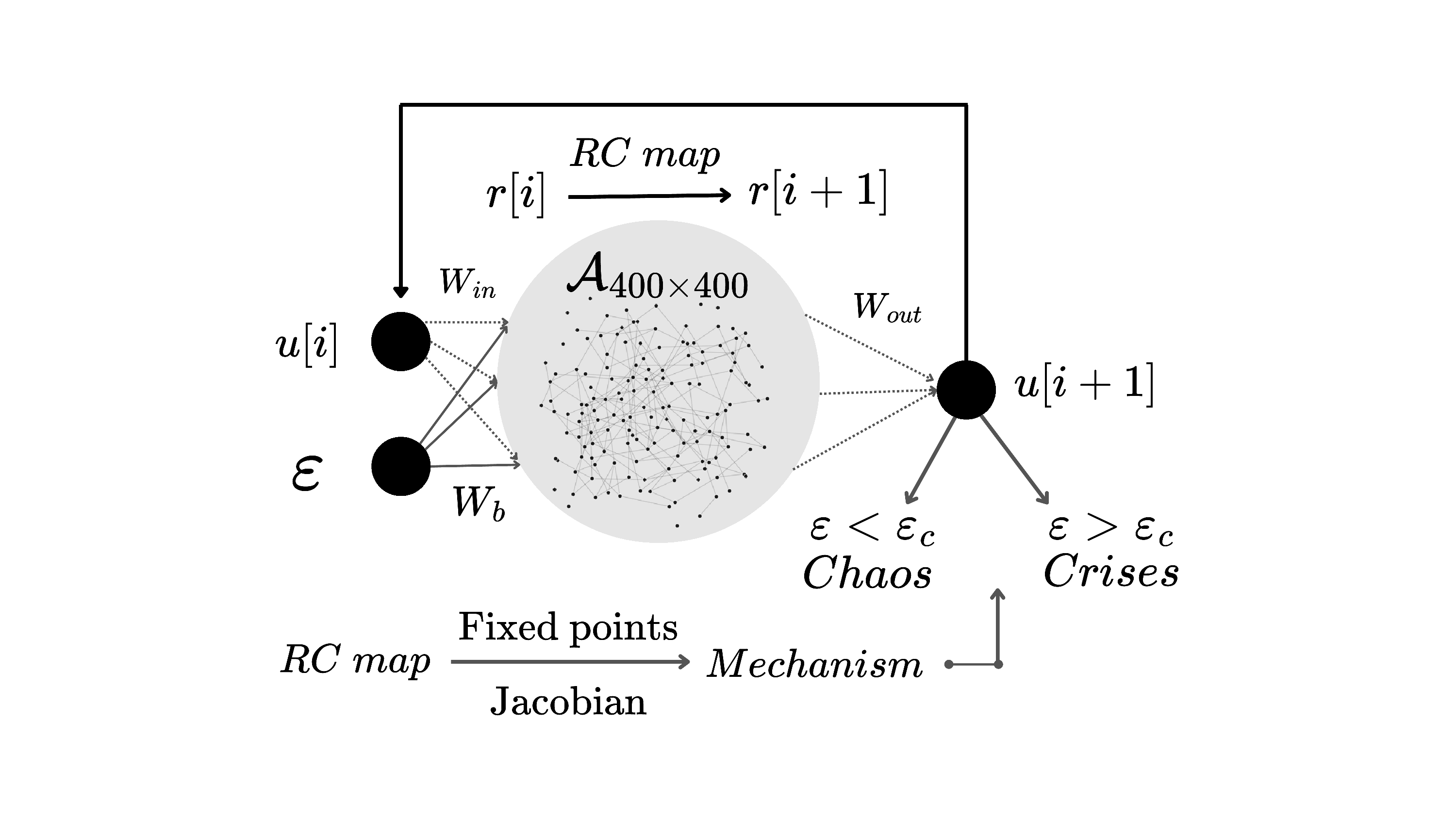}
\caption{Schematic of the trained RC map having the bifurcation parameter as an additional input channel. The weight matrices $W_{in}$ and $W_b$ map the input $u$ and $\varepsilon$, respectively, into a high dimensional reservoir space. $\mathcal{A}$ denotes the adjacency matrix of the reservoir (400 $\times$ 400 in our case), while $W_{out}$, learned during training, projects the reservoir states to give the output. Analysis of the fixed points of the trained RC map, reveals how it captures the critical transition successfully.}
\label{fig1}
\end{figure}
This Letter focuses on understanding how RC predicts crises in dynamical systems.
A phenomenon associated with sudden qualitative change of chaotic dynamics such as change in size, sudden appearance or destruction of chaotic attractor is called \textit{crisis} \cite{PhysRevLett.48.1507}. In real-world systems, slow parameter drift can induce discontinuous transitions, producing catastrophic outcomes such as tipping \cite{PhysRevLett.125.228701, PhysRevResearch.6.043194} or crises. Focusing on crises provides a pathway to uncover the mechanisms by which RC predicts such abrupt dynamical transitions.  We train the RC solely within the chaotic regime to predict the discontinuous transition without knowledge of the governing equations. Analysis of fixed points and return maps reveals the mechanism enabling accurate crisis prediction and scaling. The trained reservoir reproduces attractor collisions with unstable fixed points, mimicking the actual system, and learns the quadratic nonlinearity and invariant density responsible for the observed scaling behavior.
 
\paragraph{\bf{Crises in RC map:}}
A reservoir projects an $n$ dimensional input $u(t)$ into an $m >> n$ dimensional space via the matrix $W_{in}$, and evolves its states according to the projected input and the memory of previous states. These high-dimensional states are then mapped back to $n$ dimensions through $W_{out}$, which is optimized using ridge regression to minimize prediction error. Prior studies demonstrated that augmenting the input with the bifurcation parameter enables RC to predict critical transitions such as crises (Fig. \ref{fig1}) and amplitude death \cite{PhysRevResearch.3.013090,xiao2021predicting}.
The update rule of the reservoir states is the following:
\begin{equation}\label{eq1}
\begin{aligned}
    r[i+1] =& (1-\alpha)r[i]  +  \alpha \hspace{0.1cm} tanh(A r[i] + \\&W_{in} u[i] + k_{b} W_b (\varepsilon - \varepsilon_b)).
\end{aligned}
\end{equation}
Here $\varepsilon$ is the bifurcation parameter. $\mathcal{A}$ is chosen to be the adjacency matrix of an Erd\"os–R\'enyi model network, having a connection probability $\sigma$ and largest eigenvalue $\rho$. The matrices $W_{in}$ and $W_{b}$ are initially chosen from a uniform random distribution ranging from $[-b,b]$, $[-c, c]$ respectively and are fixed throughout the training process.
We train the reservoir at $k$ values of the bifurcation parameter prior to the critical transition. The resulting reservoir states are collected into a matrix $\mathcal{R} \in \mathbb{R}^{m \times kN_t}$, where $N_t$ denotes the number of time steps for each $\varepsilon$. The corresponding system outputs are arranged into $\mathcal{U} \in \mathbb{R}^{n \times kN_t}$. The output weights $W_{\mathrm{out}}$ are then obtained via ridge regression to minimize the prediction error;
\begin{equation}
    W_{out} = \mathcal{U}.\mathcal{R}^{T}.(\mathcal{R}\mathcal{R}^{T} + \beta \mathcal{I})^{-1}. \nonumber
\end{equation}
Once $W_{out}$ is calculated, the RC becomes an autonomous system and the trained reservoir map can be written as follows:
\begin{equation}\label{eq2} 
\begin{aligned}
    r[i+1] =& (1-\alpha)r[i]  +  \alpha \hspace{0.1cm} tanh(A r[i] + \\&W_{in}W_{out} r[i] + k_{b} W_b (\varepsilon - \varepsilon_b)),
\end{aligned}
\end{equation}

\begin{figure}[t!]
    \centering

        \includegraphics[scale=0.28]{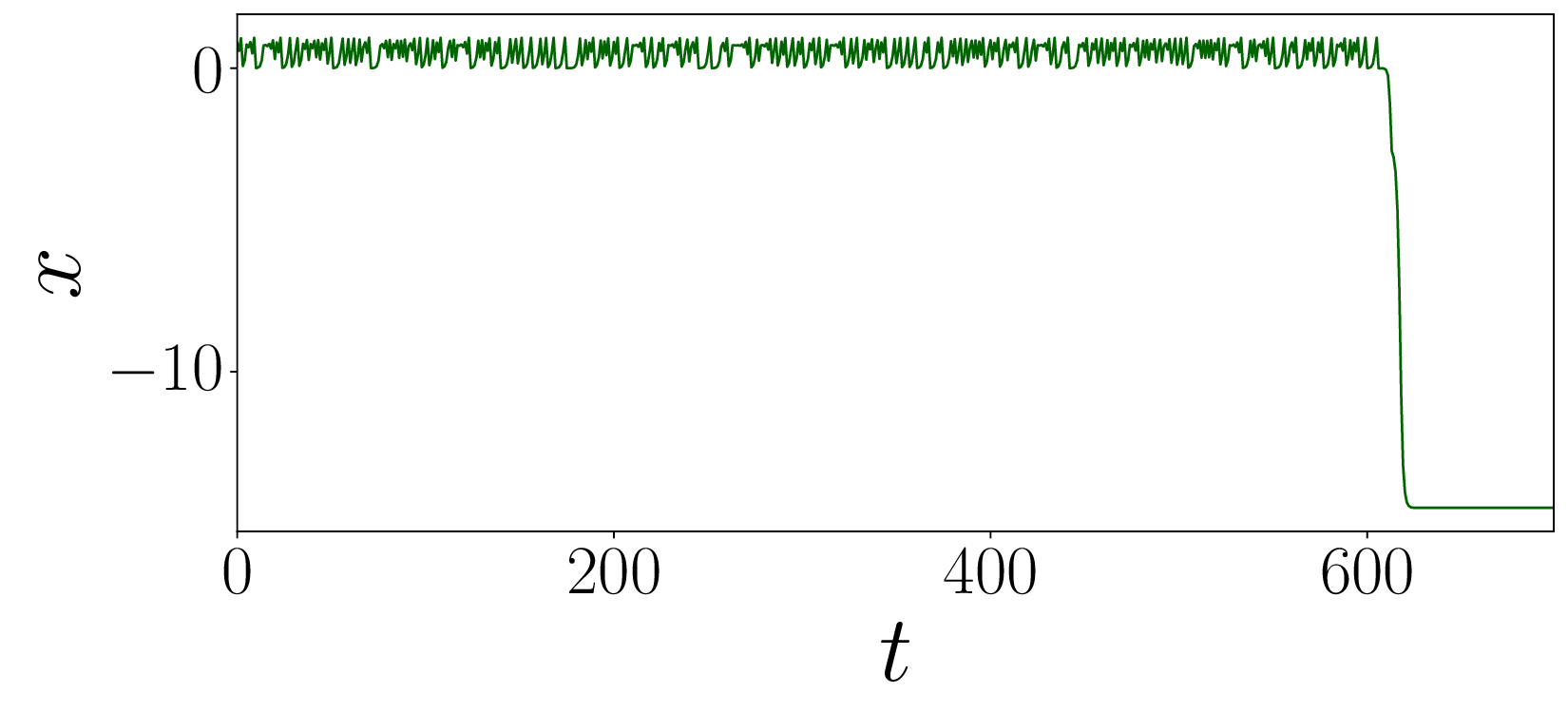}\\[2ex]
        \centering
        \includegraphics[scale=0.28]{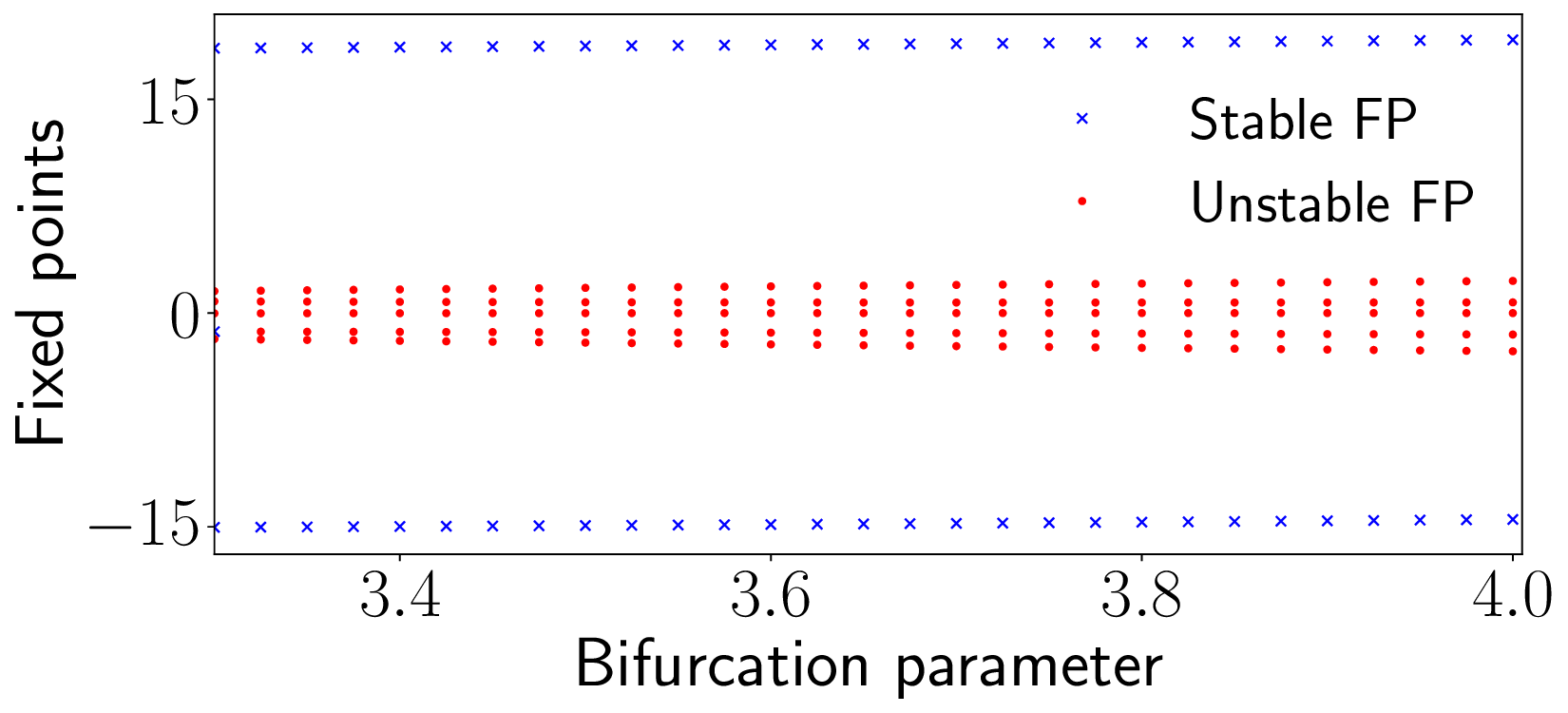}

    \caption{ (a) Time series of the trained reservoir map for $\mu_c < \mu = 3.9988$ showing the transient dynamics before convergence to a stable fixed point. (b) Fixed points (FP) of the trained RC map as the bifurcation parameter is varied. Blue crosses and red circles correspond to the stable and unstable fixed points, respectively.}
    \label{fig:2}
\end{figure}

RC is trained using the chaotic time series of the logistic map ($x_{n+1} = \mu x_n (1-x_n)$) at $\mu \in \{3.92, 3.93, 3.94, 3.95\}$. The chosen set of hyperparameters are $m = 400, \alpha = 0.86, b = 2.13, c = 1.15, \sigma = 0.526, \rho = 0.9, k_b = 1, \varepsilon_b = 0.3 $. The time series generated by the trained map (Fig. \ref{fig:2} a) predicts the transition yielding the critical transition point as $\mu_c \sim 3.9985$. We probe the trained RC map to decipher the mechanism behind this prediction by tracking the fixed points of the trained map and analyzing their stability. 

The RC architecture is a high dimensional non-linear system, therefore finding all its fixed points through traditional solvers, such as local search from random seeds, is difficult. A technique to evaluate fixed points, specially suited for RNN architecture was formulated using mathematical objects called \textit{Directional Fibers} \cite{8016349,directional-fibers} which is shown to consistently locate a large set of fixed points and is computationally competitive. A directional fiber is the inverse image of a fixed direction. For the map $f(r): \mathds{R}^{N} \rightarrow \mathds{R}^{N}$ with $f(r)=\Delta r$ denoting the change in reservoir state after one dynamical update, the fiber is the set of states where $f(r)=\alpha c$ for some constant direction $c$. It forms a one-dimensional manifold containing all fixed points, located at $\alpha=0$. The algorithm selects a random direction $c$ and traverses the corresponding fiber in state space to locate fixed points satisfying $f(r^*)=0$.

We use directional fibers approach with $f(r) = -\alpha(r - tanh(\Lambda r + \Omega (\varepsilon - \varepsilon_b)))$, where $\Lambda = A + W_{in}W_{out}$ and $\Omega = k_b W_b$ along with computing the maximum eigenvalue of the Jacobian of reservoir map at the fixed point to check for the stability. For an analytical expression of Jacobian, we can introduce a variation:
\begin{equation} 
\begin{aligned}
    \delta r[i+1] &= (1-\alpha)\delta r[i] + \{\mathbb{I} - tanh^{2}(\Lambda r^{*} + \Omega)\}\Lambda \delta r[i] \\&
    = \mathcal{J} \delta r[i].    \nonumber
\end{aligned}
\end{equation}
We obtain seven fixed points as illustrated in (Fig. \ref{fig:2} b), out of which two are stable and rest are unstable. The stable fixed points are the only attractors in the system after $\mu = 4$, hence the trajectories can only settle there instead of diverging off to $-\infty$ as it happens in the original logistic map.

\begin{figure}[t!]
    \centering

        \includegraphics[scale=0.24]{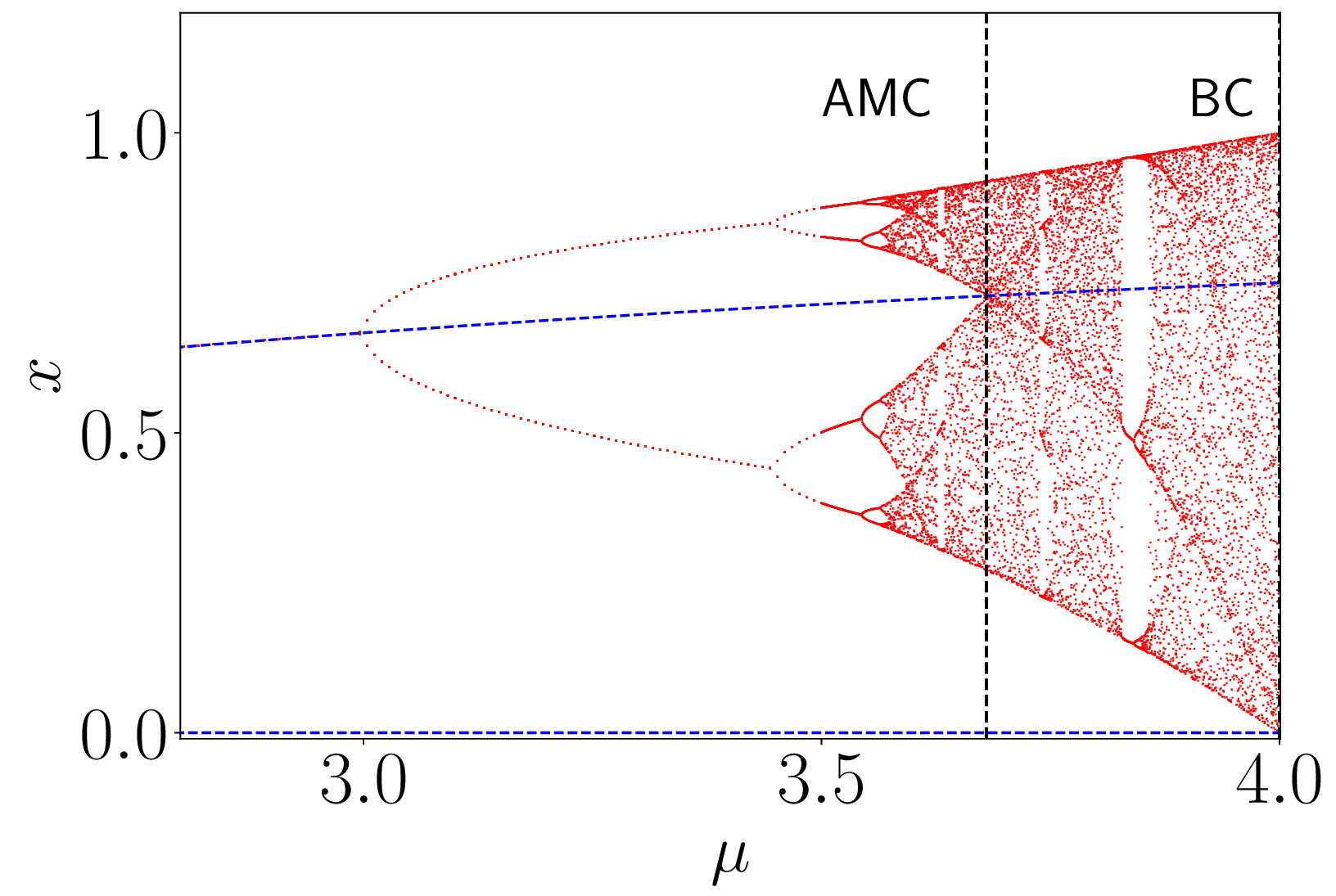} \\[2ex]
        \centering
        \includegraphics[scale=0.24]{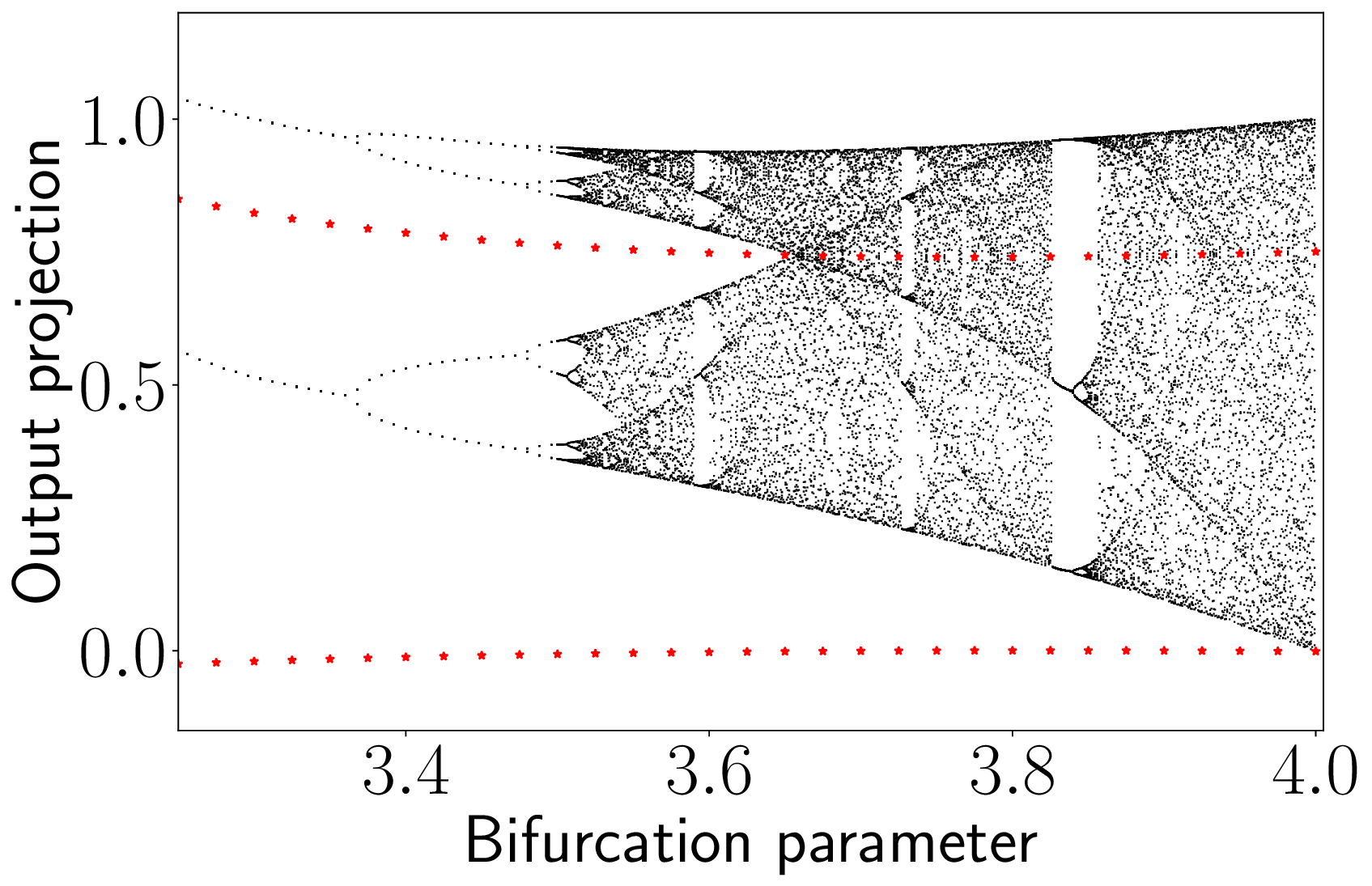}

    \caption{ (a) Crises in a logistic map. The dashed blue lines show two fixed points $x^* = 0$ and $x^* = 1 - 1/\mu$. The first vertical dashed red line at $\mu = 3.68$ corresponds to attractor merging crises (AMC). The second vertical dashed line at $\mu = 4$ shows boundary crisis (BC). (b) Crises in trained RC map due to collision of two unstable fixed points with the chaotic attractor. Collision close to $\mu = 3.68$ is responsible for attractor merging crisis and the collision close to $\mu = 4$ causes boundary crisis.}
    \label{fig:3}
\end{figure}

A boundary crisis is the abrupt destruction of a chaotic attractor and its basin as a bifurcation parameter is varied \cite{PhysRevLett.48.1507}. It occurs when the stable manifold of an unstable periodic orbit forming the basin boundary collides with the chaotic attractor lying on its unstable manifold. For parameter values past the crisis point, the trajectories initialized in the region formerly occupied by the chaotic attractor wander in that region for some finite time (chaotic transient), after which they settle on to some other attractor. For logistic map $x_{n+1} = \mu x_n (1-x_n)$, at $\mu = 4$, the chaotic attractor collides with the unstable fixed point $x = 0$ (Fig. \ref{fig:3} a) beyond which all the trajectories initialized at different initial conditions eventually diverge to $-\infty$. Besides boundary crisis, chaos can disappear through other mechanisms such as Pomeau- Manneville scanario \cite{RevModPhys.53.643}, or the unstable–unstable pair bifurcation \cite{Grebogi1985}. Thus, while the reservoir map learns to predict crises, the mechanism by which it does so remains nontrivial.

The plot of the bifurcation diagram of the trained map (Fig.~\ref{fig:3} b) along with all the evaluated fixed points is constructed by projecting Eq.~\ref{eq2} back into one-dimension using $W_{out}$ matrix. All the fixed points of the high dimensional reservoir space are also projected back to single dimension. For $\mu \sim 4$, we see collision of an unstable fixed point with the chaotic attractor after which the chaotic attractor disappears. This unstable fixed point tries to mimic $x^* = 0$, at values of $\mu$ close to $\mu = 4$. This confirms the boundary crisis route in the reservoir map. 

Attractor merging crisis (AMC) is the merging of a multi-piece chaotic attractor to increase in size. For AMC,  merging of attractors can be seen in the interval $3.57 < \mu <3.68$ (Fig.~\ref{fig:3} a). Starting from the onset of chaos ($\mu$ = 3.57), at each such crisis the number of pieces of chaotic attractor is halved \cite{ananthkrishnan2001crises}. The most prominent attractor merging crisis happens at $\mu = 3.68$ where two pieces merge to yield a single piece chaotic attractor (Fig.~\ref{fig:3} a). At each period doubling bifurcation, an unstable periodic orbit is created. These orbits collide with the pieces of chaotic attractor causing a cascade of ``halving" of the number of pieces of the attractor. The final crisis point is the collision of unstable periodic orbit of period 1 (fixed point: $x^* = 1- 1/\mu$) with the two pieces of chaotic attractor.

In the RC map, for $\mu = 3.64$, two pieces of chaotic attractor merge to form a single attractor and this precisely coincides with an unstable fixed point (unstable periodic orbit of period 1) colliding with the two pieces (Fig.~\ref{fig:3} b). The unstable fixed point continues to exist for larger values of $\mu$. Moreover, this tries to mimic $x^* = 1-1/\mu$. This confirms that the trained reservoir map learns to predict attractor merging crisis in its 1-D projection.

\paragraph{\bf{Scaling exponents near crisis in the original and RC map:}}
For boundary crisis, the average length of chaotic transient $\langle\tau\rangle$ scales as: $\langle\tau\rangle = (\mu - \mu_c)^{-\gamma}$ where $\mu_c$ is the bifurcation point and $\gamma$ is the scaling exponent. For the logistic map, $\gamma = 1/2$. The value of scaling exponents for one-dimensional maps of the form $x_{n+1} = f(x_n,\mu)$ can be obtained analytically as following. Let $x_e$, be the escape point such that $f(x_e, \mu_c) = X_{boundary}$, where $X_{boundary}$ is the boundary point of chaotic interval after which the future iterates rapidly move towards some other attractor. For logistic map: $x_{e} = 0.5$, $X_{boundary} = 1$ because beyond $\mu_c = 4$, $x > X_{boundary}$ leads to future iterates having negative values diverging off to $-\infty$.

To derive the scaling exponent \cite{Ott2002} let $\mu = \mu_c + \epsilon$, $\delta x = x - x_e$. Taylor expansion results in:
\begin{equation}
    \begin{aligned}
        f(x_e + \delta x, \mu_c + \epsilon) =& f(x_e, \mu_c) \;+\;  \frac{\partial f}{\partial x}\delta x \; + \; \frac{\partial f}{\partial \mu} \epsilon \\&+ \frac{1}{2} \frac{\partial^2 f}{\partial x^2} (\delta x)^2 + higher \; order. \nonumber
    \end{aligned}
\end{equation}
All the partial derivatives are evaluated at $x_e, \mu_c$. Since $f(x_e, \mu_c) = X_{boundary}$ and for escape $f(x, \mu) > X_{boundary}$, the escape condition becomes the following:
\begin{equation}\label{eq3}
    \begin{aligned}
        \frac{\partial f}{\partial x}\delta x \; + \; \frac{\partial f}{\partial \mu} \epsilon \;+\; \frac{1}{2} \frac{\partial^2 f}{\partial x^2} (\delta x)^2 > 0
    \end{aligned}
\end{equation}
We rewrite it as $A\epsilon + B (\delta x)^2 + C \delta x > 0$. In case of the logistic map, due to the quadratic non-linearity and the maxima occurring precisely at $x_e$, $B<0$ and $C$ becomes zero. This makes the escape condition as $(\delta x)^2 < -A\epsilon / B $ or $\delta x \sim \sqrt{\epsilon}$, i.e. if the trajectory lands in the region of width $\delta x$, the future iterates diverge to $-\infty$ ending the transient length. The invariant density $\rho(x) dx$ is defined as the probability that a point on a long trajectory will lie between $x$ and $x + dx$. To compute it numerically, we make a histogram of the iterates of any initial condition on the attractor. Since the natural invariant density is uniform for logistic map, the probability per iterate of an orbit falling in the small region $\delta x$ is proportional to $\delta x$ or $\sqrt{\epsilon}$. The transient length is inversely proportional to this probability, therefore $\langle\tau\rangle = (\mu - \mu_c)^{-1/2}$. 

It has been shown that a trained reservoir can mimic statistical properties such as the scaling exponent \cite{Kong2021}. By taking the average over different stochastic realizations of reservoir one can compute the scaling exponent, however it changes the high dimensional dynamical system each time due to randomness involved in the initialization of the weight matrices.
This approach does not allow the analysis of the RC map to understand how the scaling exponent matches the true system. Therefore, we use a different approach for scaling exponent calculation and explain the mechanism behind its successful prediction. For a generalized one-dimensional map $x_{n+1} = f(x_n, \mu)$, the escape condition is given by Eq.~\ref{eq3}. For the case of trained reservoir map we have:
\begin{equation}\label{eq4} 
\begin{aligned}
    x_{i+1} =& W_{out} \Big ((1-\alpha) r_{i} \\&+  \alpha \;tanh(Ar_{i} + W_{in}x_{i} + \Omega(\varepsilon - \varepsilon_b)) \Big).
\end{aligned}
\end{equation}
To calculate the scaling exponent, we randomly sample initial condition $x_0$ from the interval $(0.2, 0.6)$. Using the logistic map equation, future iterates are generated for reservoir warm up. The reservoir state is initialized as an array of zeros, but is updated using Eq.~\ref{eq1} for some iterates, commonly known as the \textit{warm up} of the states in RC literature. Then using this warm up state as initial condition, for a particular value of bifurcation parameter, RC is updated as an autonomous system (Eq.~\ref{eq2}) and $W_{out} r[i+1] = u[i+1]$. It keeps on updating till $u[i+1] < 0$. The number of steps taken to hit this threshold is stored for every initial condition and the average number of steps is $\langle\tau\rangle$.

\begin{figure}[t!] 
    \includegraphics[scale=0.22]{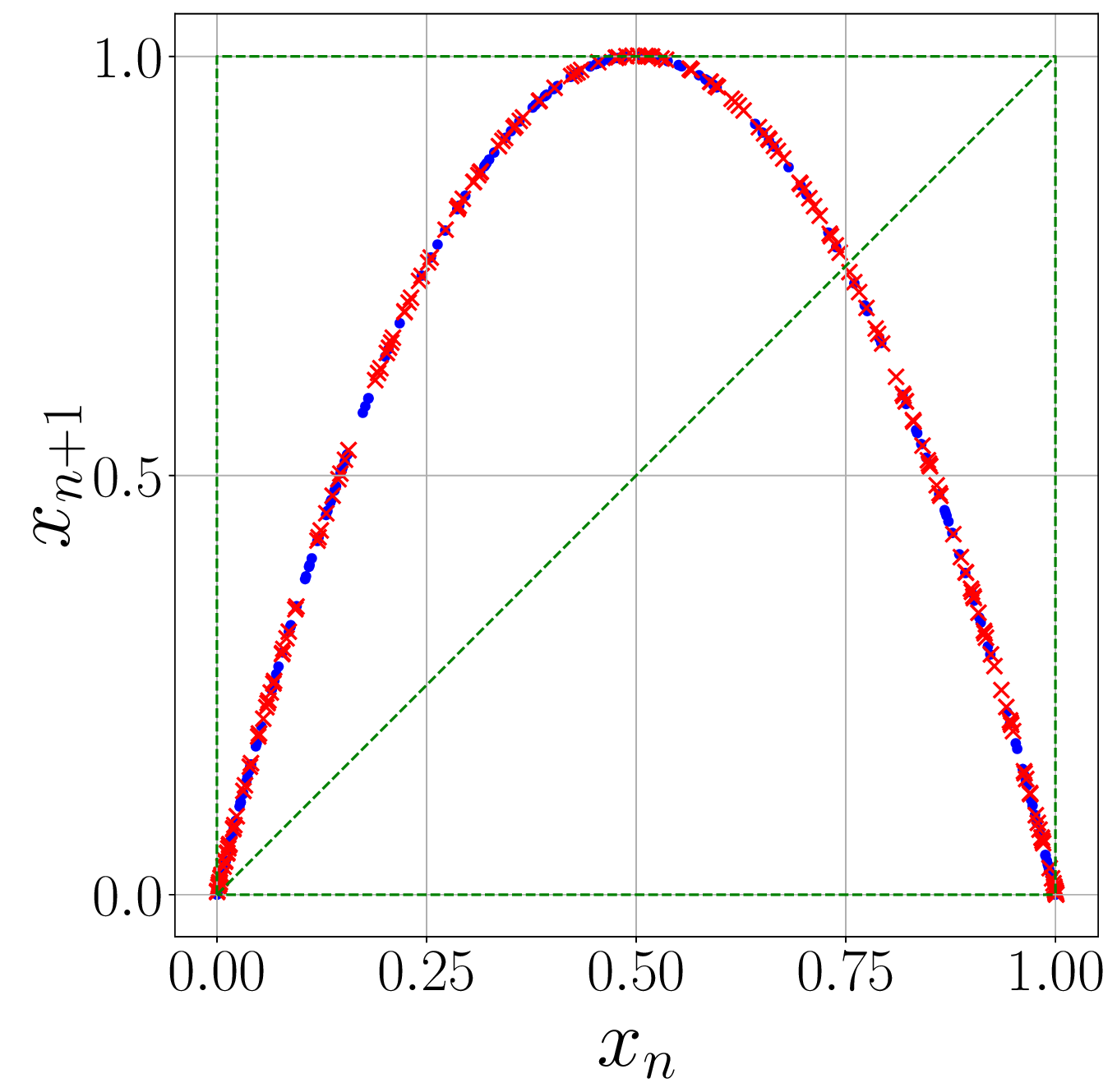}
    \caption{Return map of trained reservoir map (red crosses) along with the logistic map (blue dots). The green box denotes the bounded chaotic region, outside which the trajectories escape.}
    \label{fig4}
\end{figure}

In the transient regime, the return map $(x_{i+1}\; vs \; x_{i}) $ of the trained reservoir and the logistic map (Fig.~\ref{fig4}) at $\mu$ and $\varepsilon$ (such that $\mu - \mu_c = \varepsilon- \varepsilon^* = 0.0001$) show the same parabolic nature with one maxima at the escape point $x = 0.5$. Therefore, the escape condition for the trained reservoir map may also be given by $A \epsilon +B (\delta x)^2 > 0$.

\begin{figure}[t!]
    \includegraphics[scale=0.25]{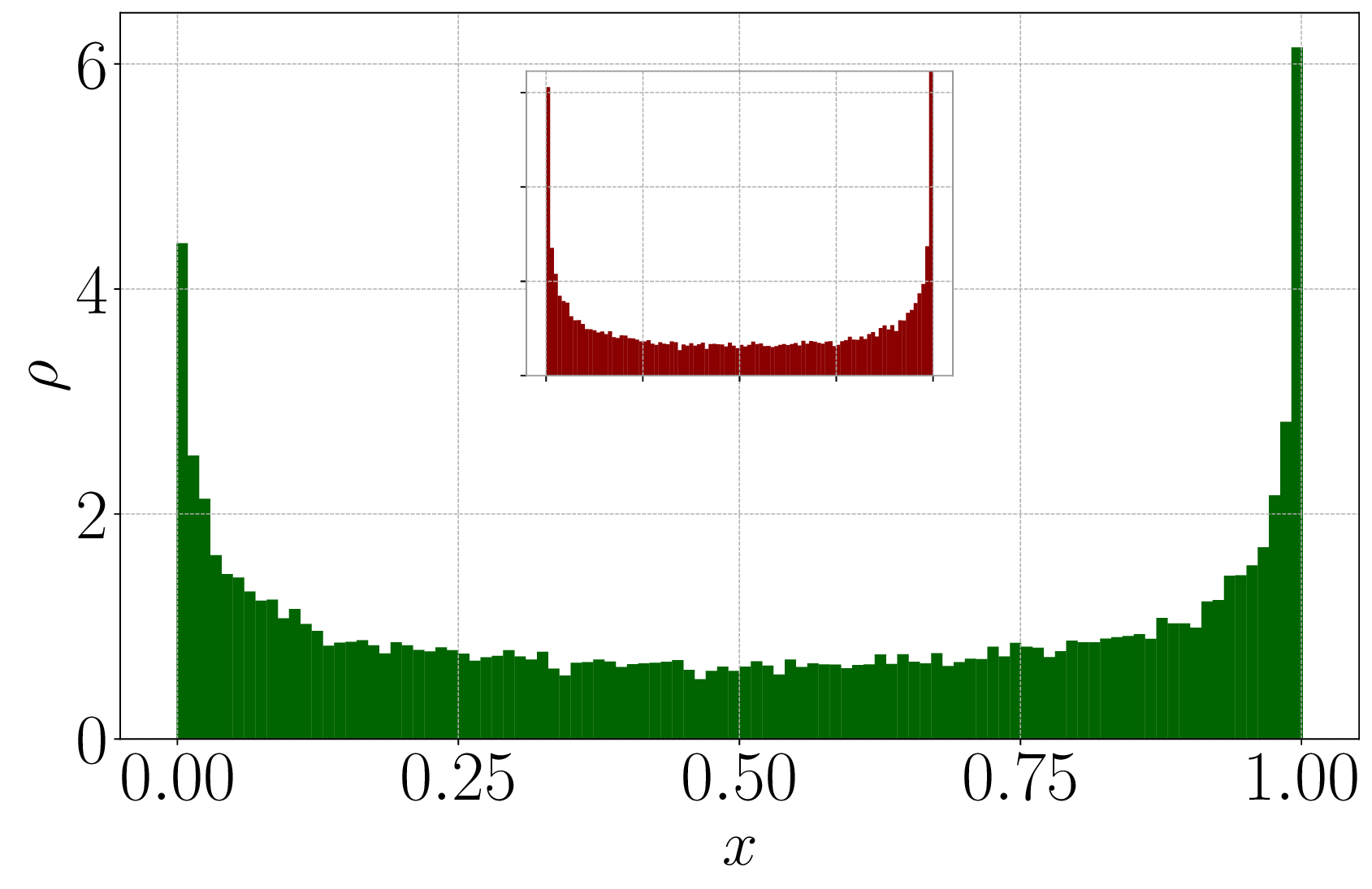}
    \caption{Histogram of 50000 iterates of reservoir map shows a uniform $\rho(x)$ similar to logistic map (shown in the inset).}
    \label{fig5}
\end{figure}

\begin{figure}[t!]
     \includegraphics[scale=0.3]{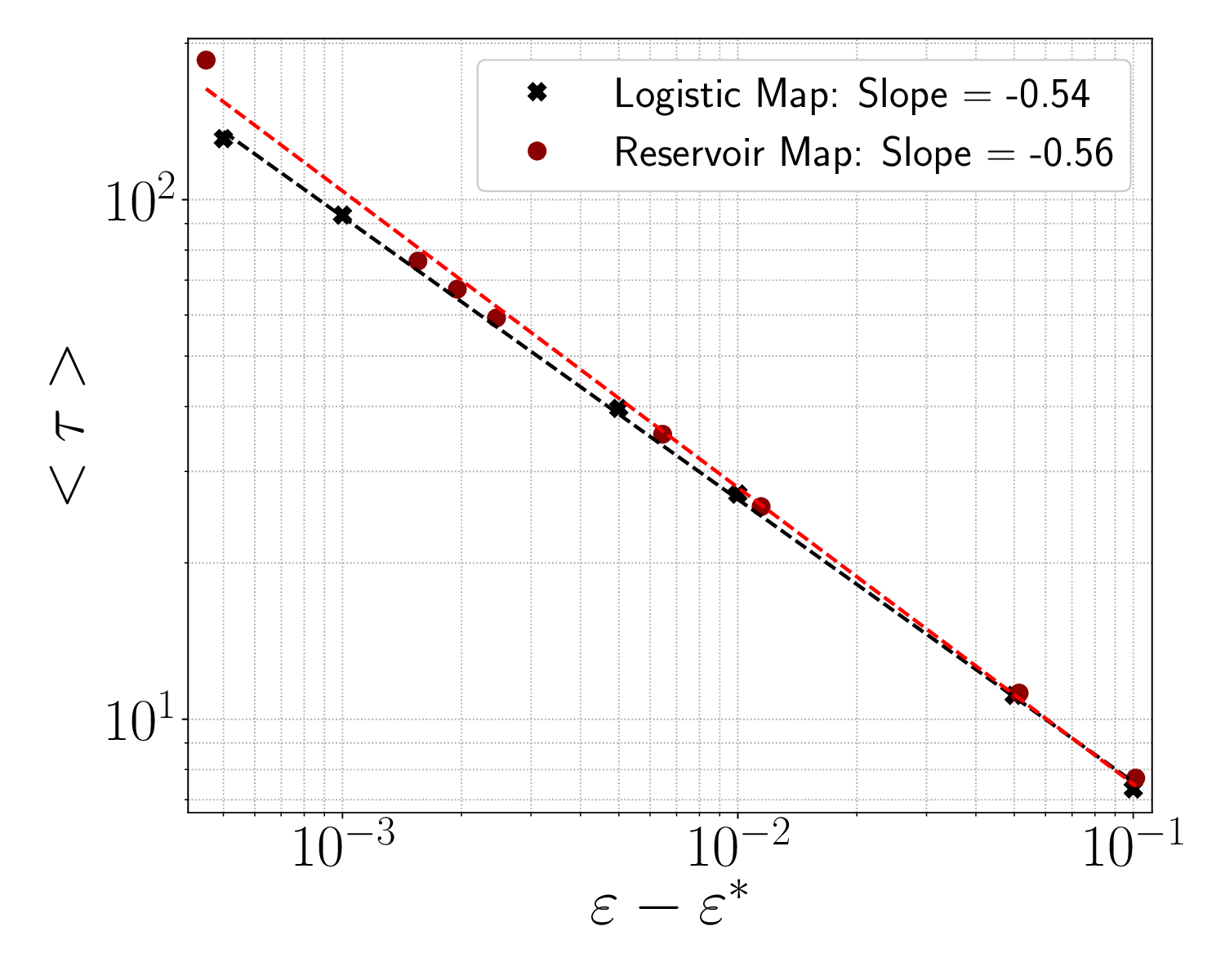}
    \caption{$\langle \tau \rangle$ is plotted as a function of $\varepsilon - \varepsilon^*$. Trained reservoir map matches the scaling exponent of the logistic map. The average over $\tau$ is performed using 10000 different initial conditions.}
    \label{fig6}
\end{figure}

Further we compute the natural invariant density of the trained reservoir map (Fig.~\ref{fig5}) which is done by picking up a random initial condition $(x_0)$ and warming up the reservoir state. The reservoir state obtained after the \textit{warm up} is used as an initial condition and the iterates multiplied by $W_{out}$ are used to plot the histogram. It closely resembles the invariant density of the logistic map. Therefore, the probability to fall in a small region $\delta x$ is proportional to $\delta x$ and the transient length being inversely proportional to the probability yields $\langle\tau\rangle = (\mu - \mu_c)^{-1/2}$. This explains how the trained reservoir is able to correctly infer the scaling exponent (Fig.~\ref{fig6}) of the original logistic map system.

Due to the wide range of tunable hyperparameters ($m, \alpha, b, c, \sigma, \rho, k_b, \varepsilon_b$), the learned reservoir map can display significant variations through changes in the underlying matrices while preserving its functional form. While this Letter reports results for a particular choice of hyperparameters, other sets can yield reservoir maps with different matrix realizations that still predict crises successfully. One such variation, presented in the Supplementary Material \cite{supp}, confirms that the mechanism of attractor disappearance remains a boundary crisis, characterized by an algebraic scaling law. For example, training the RC on logistic map time series at widely spaced parameter values $\mu \in \{3.8, 3.85, 3.9, 3.95\}$, with all other hyperparameters fixed, produces a bifurcation diagram that differs from the case with closely spaced $\mu$ values. A stable period-2 orbit emerges after the critical bifurcation point. Nevertheless, the crisis mechanism is preserved as a boundary crisis with algebraic scaling.

\paragraph{\bf{1-d Gauss map:}}
To demonstrate the generality of our approach, we train the RC on another one-dimensional map called Gauss map \cite{Oliveira2022}, and investigate the dynamics of trained reservoir map. Gauss map is characterized by two control parameters ($\nu$ and $\beta$):
\begin{equation}
    x_{n+1} = exp(-\nu x_n^2) + \beta.    \nonumber
\end{equation}
We fix $\nu = 10$ and vary $\beta$ as the bifurcation parameter. An attractor–merging crisis (AMC) occurs near $\beta \sim -0.82$, while a boundary crisis (BC) arises near $\beta \sim -0.682$. Following the analysis of the RC map trained on the logistic map, we investigate the mechanism underlying successful crisis prediction and the accurate recovery of the scaling exponent (details in the Supplementary material \cite{supp}.

\paragraph{\bf{Conclusion:}}
Using the logistic map and Gauss map as representative examples, we showed how a trained reservoir map can capture subtle transitions that occur in chaotic dynamics. The results establish that the reservoir does more than just reproducing the trajectories, infact, it successfully replicates crises phenomena, namely, boundary and attractor-merging crises. Moreover, it recovers the correct scaling behavior of transient lifetimes near the crisis point. We demonstrate that the trained reservoir map also displays collision of chaotic attractor with unstable fixed points, a behavior characteristic of the model system on which RC is trained, despite the existence of alternative mechanisms \cite{RevModPhys.53.643,Grebogi1985} for attractor disappearance. Further, the RC map analysis reveals how the correct scaling exponent can be predicted. We further examine the challenges of crisis prediction in higher-dimensional systems in the supplementary material \cite{supp}, using the Hénon map \cite{PhysRevE.49.1094} as an example. Although the RC map successfully reproduces the crisis, its high dimensionality makes it virtually impossible to visualize the stable and unstable manifolds and to pinpoint the underlying mechanism. These difficulties become even more severe in higher-dimensional models, such as the three-dimensional food chain system \cite{mccann1994}, where boundary crisis are governed by saddle-type periodic orbits. In such cases, one must both identify a periodic orbit of unknown period within the RC map and track its manifolds to verify their intersection at the crisis. Developing strategies to overcome these challenges will be an important direction for future work, enabling deeper insight into how data-driven models can capture the mechanisms of discontinuous transitions.

\begin{acknowledgments}
SJ gratefully acknowledges SERB Power grant award SPF/2021/000136. DS thanks CSIR fellowship under award no. JRF-NET2024/14347. SJ thanks M. Small for constructive suggestions.
\end{acknowledgments}

\twocolumngrid

\clearpage

\onecolumngrid
\clearpage
\renewcommand\theequation{S\arabic{equation}}
\renewcommand\thefigure{S\arabic{figure}}
\setcounter{equation}{0}
\setcounter{figure}{0}

\begin{center}
    {\bf \large Supplemental Material for ``Dynamics of reservoir computing for crises prediction''}
\end{center}
\vspace{1em}

\noindent
\textbf{Authors:} Dishant Sisodia and Sarika Jalan\\
\textit{Complex Systems Lab, Department of Physics, Indian Institute of Technology Indore, Khandwa Road, Simrol, Indore-453552, India}

\vspace{1em}
\hrule
\vspace{1em}

\section{Training RC on Gauss map}
\begin{figure}[h!]
    \centering
    \begin{minipage}{0.48\linewidth}
        \centering
        \includegraphics[scale=0.27]{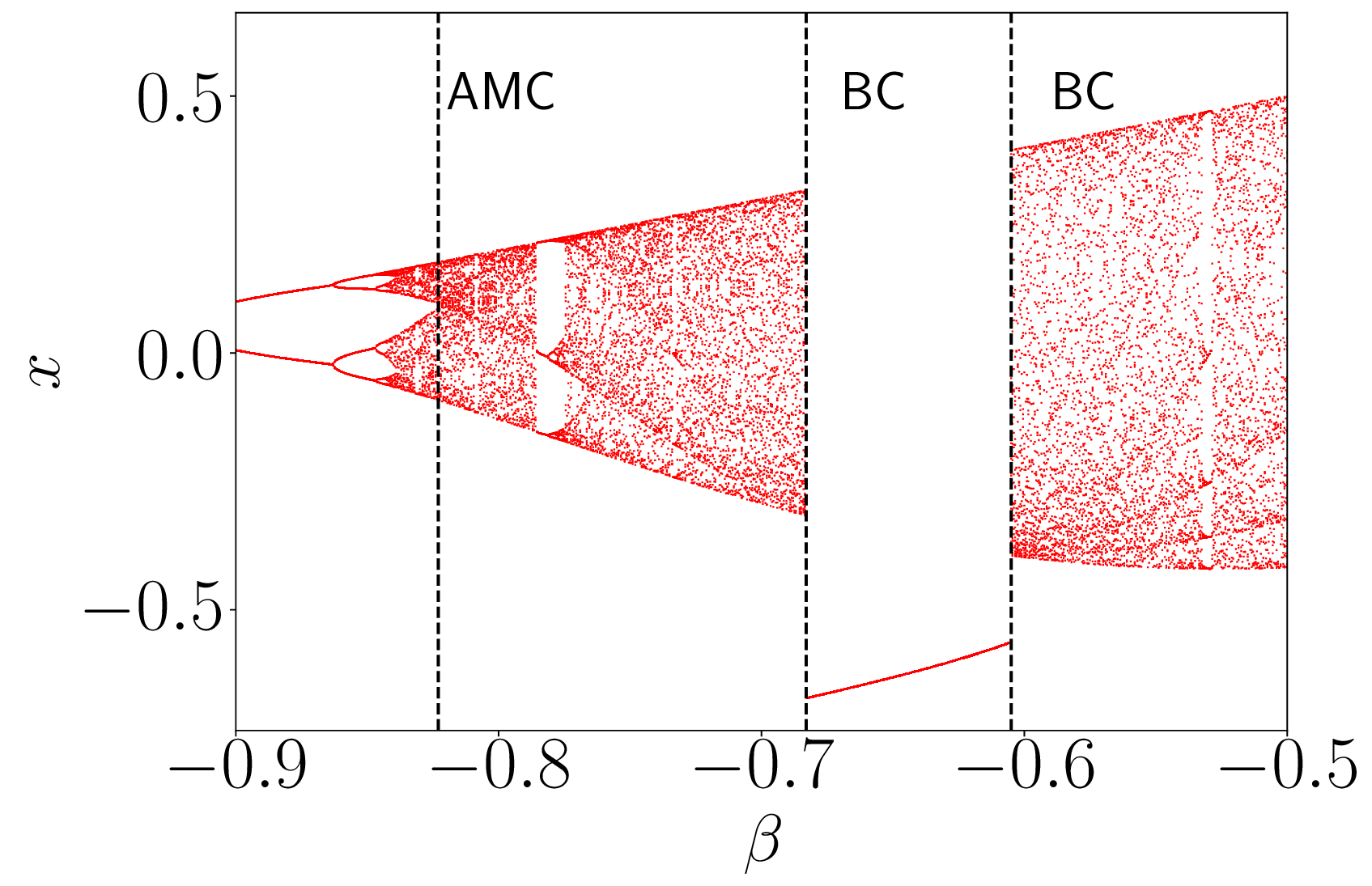}
        \caption{Crises in a gauss map. The first and second vertical dashed black lines at $\beta \sim -0.82$ and  $\beta \sim -0.682$ corresponds to AMC and BC, respectively. The third dashed black line at $\beta \sim -0.6$ shows the reappearance of chaotic attractor through inverse boundary crisis route.}
        \label{figs1}
    \end{minipage}
    \hfill
    \begin{minipage}{0.48\linewidth}
        \centering
        \includegraphics[scale=0.28]{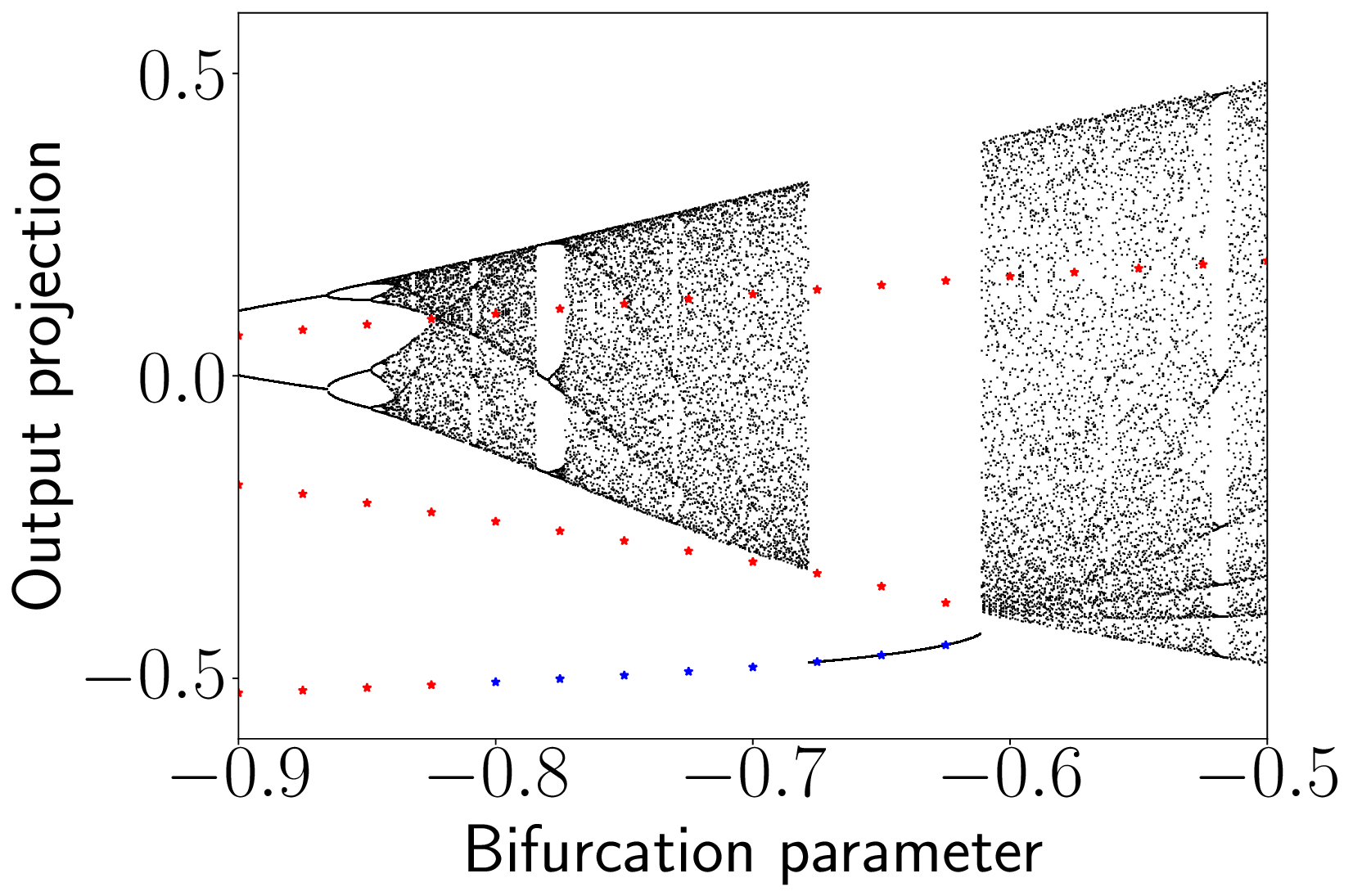}
        \caption{Crises in trained RC map due to collision of two unstable fixed points with the chaotic attractor. Collision close to $\beta \sim - 0.82$ is responsible for AMC and the collision close to $\beta \sim -0.68$ causes BC.}
        \label{figs2}
    \end{minipage}
\end{figure}

To demonstrate the generality of our approach to elucidate the mechanism for crises prediction, we train the RC on another one-dimensional map called Gauss map, and investigate the dynamics of trained reservoir map. Gauss map \cite{Oliveira2022} is characterized by two control parameters ($\nu$ and $\beta$) as;
\begin{equation}
    x_{n+1} = exp(-\nu x_n^2) + \beta.    \nonumber
\end{equation}
We fix $\nu = 10$ and treat $\beta$ as the bifurcation parameter. Upon increasing $\beta$, period doubling route to chaos is observed (Fig. \ref{figs1}). When $\beta \sim -0.82$, attractor merging crisis (AMC) is observed and for $\beta \sim -0.682$, boundary crisis (BC) is observed. Another peculiar feature of this map is an existence of inverse boundary crisis at $\beta \sim -0.6$ where the chaotic attractor reappears as $\beta$ increases. The RC map is trained in the chaotic regime at 6 different values of $\beta \in \{-0.76, -0.745, -0.73, -0.715, -0.7, -0.685\}$. The chosen set of hyperparameters are $m = 400, \alpha = 0.86, b = 2.13, c = 1.15, \sigma = 0.526, \rho = 0.9, k_b = 1, \varepsilon_b = 0.3$. 

\begin{figure}[h!]
    \centering
    \begin{minipage}{0.48\linewidth}
        \centering
        \includegraphics[scale=0.28]{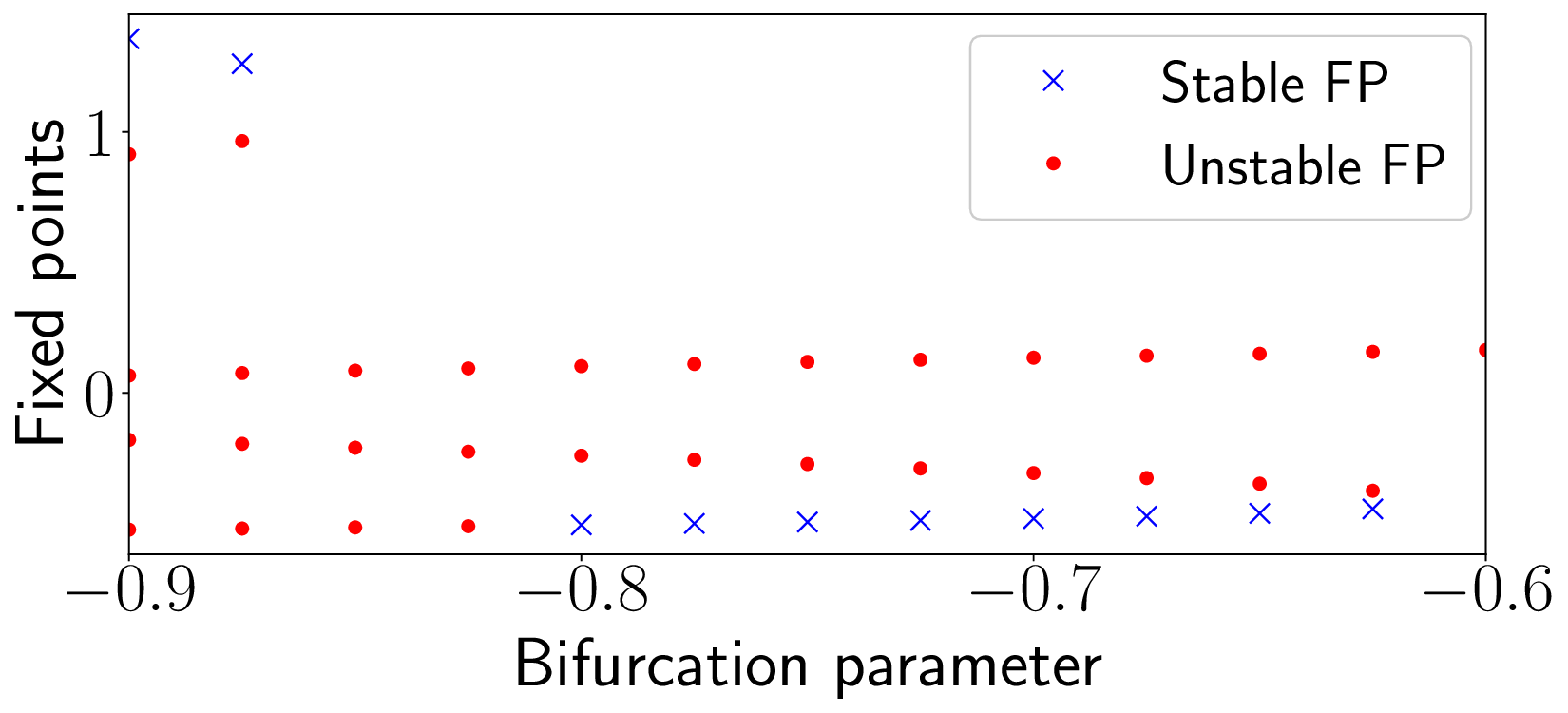}
        \caption{Fixed points of the trained RC map as the bifurcation parameter is varied. Blue crosses and red circles correspond to the stable and unstable fixed points, respectively.}
        \label{figs3}
    \end{minipage}
    \hfill
    \begin{minipage}{0.48\linewidth}
        \centering
        \includegraphics[scale=0.26]{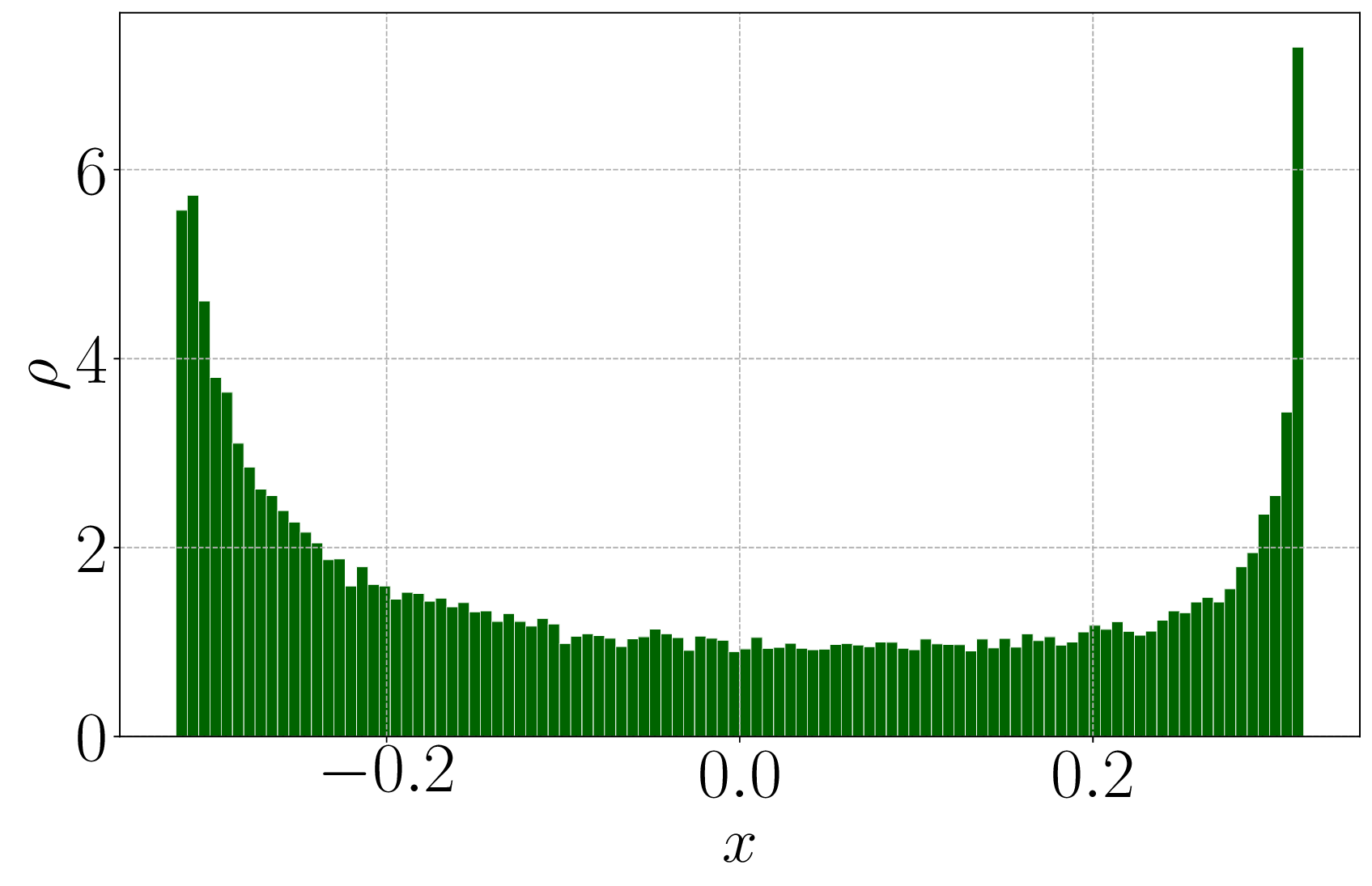}
        \caption{Histogram of 50000 iterates of reservoir map shows a uniform $\rho(x)$ similar to gauss map.}
        \label{figs4}
    \end{minipage}
\end{figure}

\begin{figure}[h!]
    \centering
    \begin{minipage}{0.48\linewidth}
        \centering
        \includegraphics[scale=0.27]{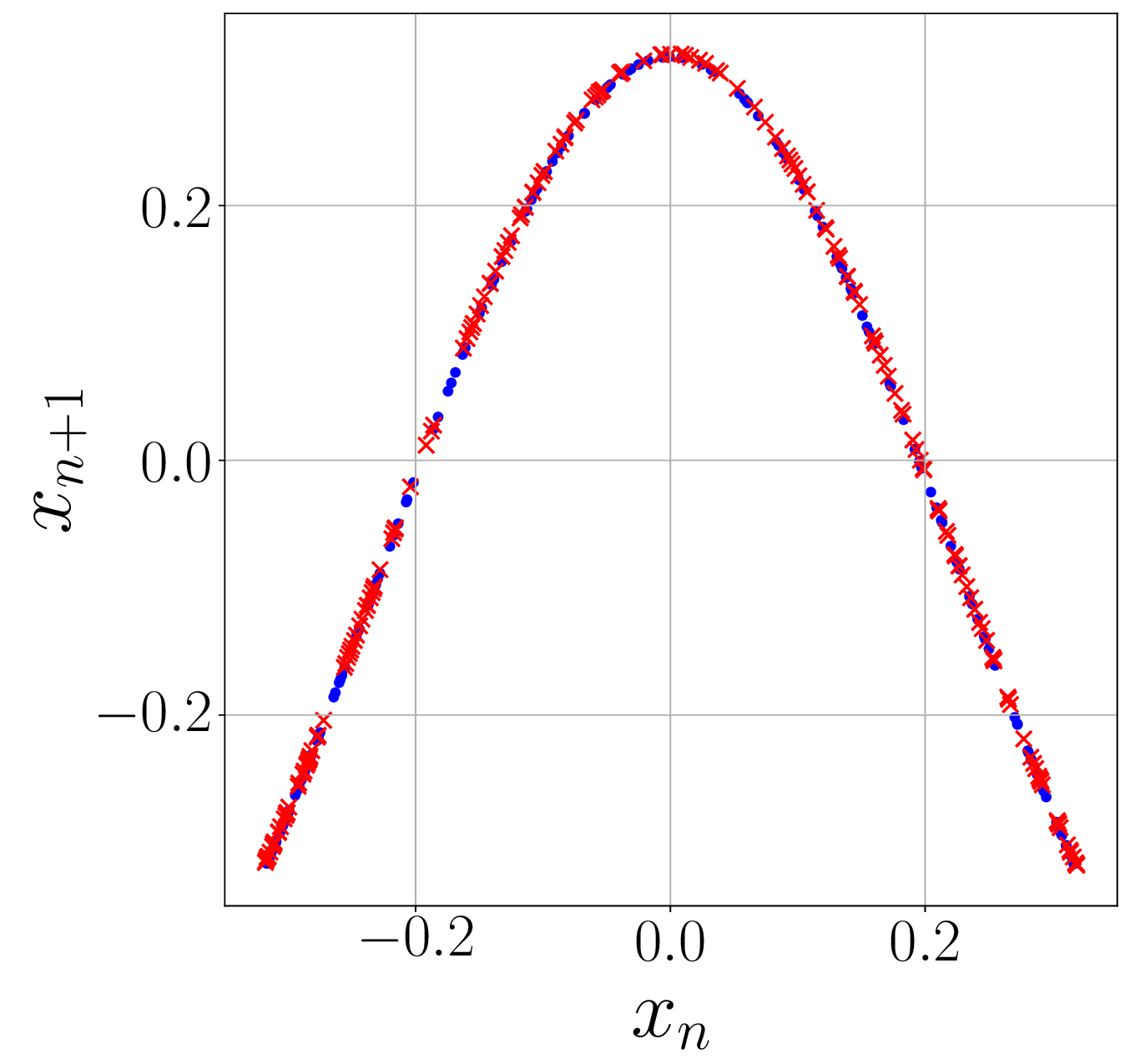}
        \caption{Return map of trained reservoir map (red crosses) along with the gauss map (blue dots) shows the same quadratic nature with a maxima.}
        \label{figs5}
    \end{minipage}
    \hfill
    \begin{minipage}{0.48\linewidth}
        \centering
        \includegraphics[scale=0.3]{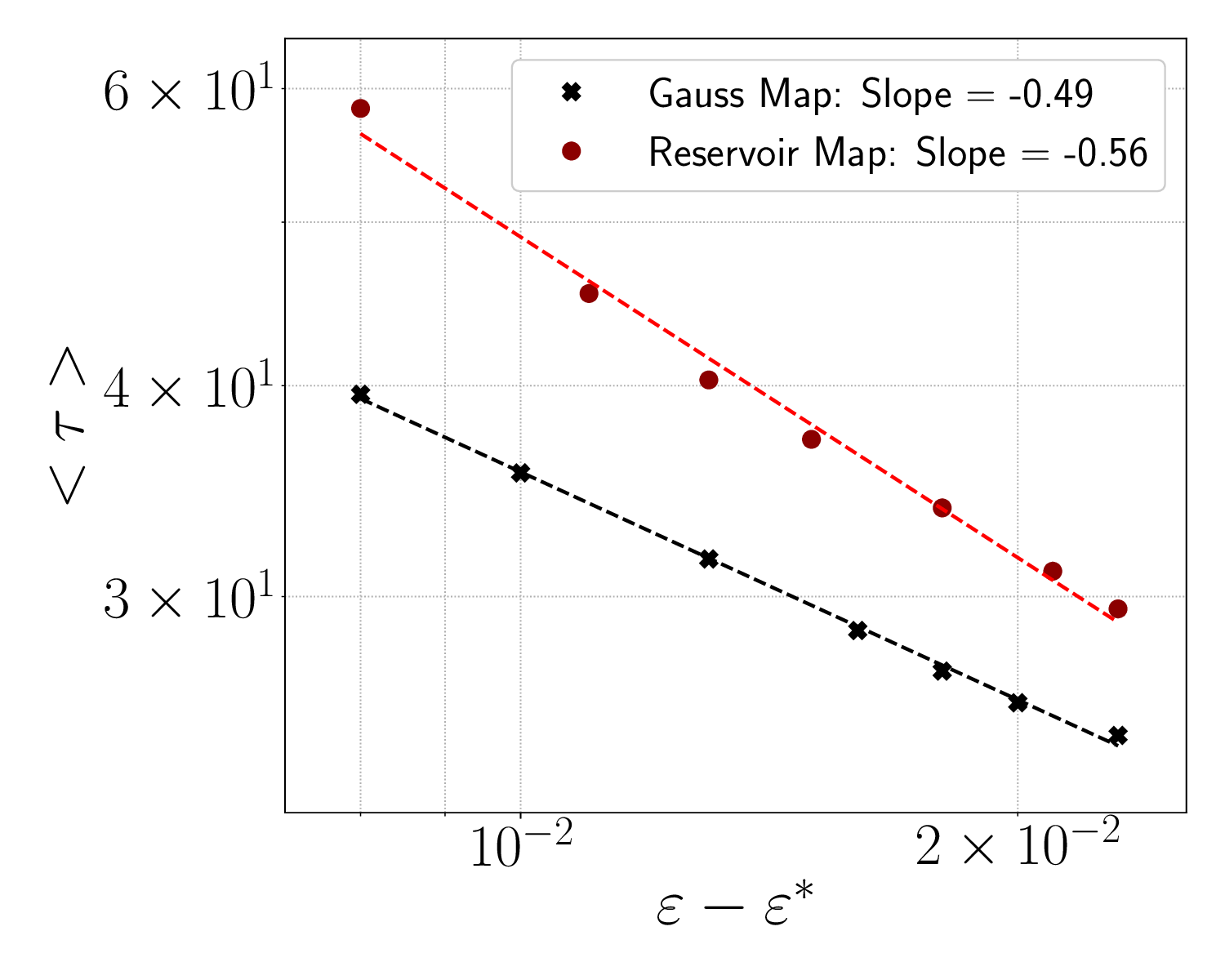}
        \caption{$\langle \tau \rangle$ is plotted as a function of $\varepsilon - \varepsilon^*$. Trained reservoir map matches the scaling exponent of the gauss map. The average over $\tau$ is performed using 10000 different initial conditions.}
        \label{figs6}
    \end{minipage}
\end{figure}

Fig.~\ref{figs2} plots the bifurcation diagram of the trained map along with all the evaluated fixed points by projecting them back into one-dimension using $W_{out}$ matrix.
After training, the fixed points of the trained RC map along with their stability were tracked through the directional fibers approach (Fig.~\ref{figs3}) similar to the logistic map case. 
Trained RC map is able to successfully reproduce both AMC and BC through the collisions with the respective unstable fixed points. It is also able to capture the inverse boundary crisis route through which the chaotic attractor reappears. Additionally, we investigate the scaling exponents near the boundary crisis ($\beta \sim -0.682$) in the original and the RC map. The natural invariant density of the trained RC map (Fig. \ref{figs4}) is uniform. The return maps (Fig. \ref{figs5}) reveal the same quadratic nature with a single maxima. 
This makes the escape condition as $(\delta x)^2 < -A\epsilon / B $ or $\delta x \sim \sqrt{\epsilon}$, i.e., if the trajectory lands in the region of width $\delta x$, the future iterates leave the chaotic attractor, ending the transient length. Therefore, the analytical calculation of scaling exponent matches that of the logistic map and the scaling exponent is thereby expected to be close to $0.5$. We calculate the scaling exponent for the trained RC map using the same method described in the Letter and the numerical calculation closely matches the analytical predictions (Fig.~\ref{figs6}).

\section{Training RC on logistic map with a different training dataset}
\begin{figure}
    \includegraphics[scale=0.3]{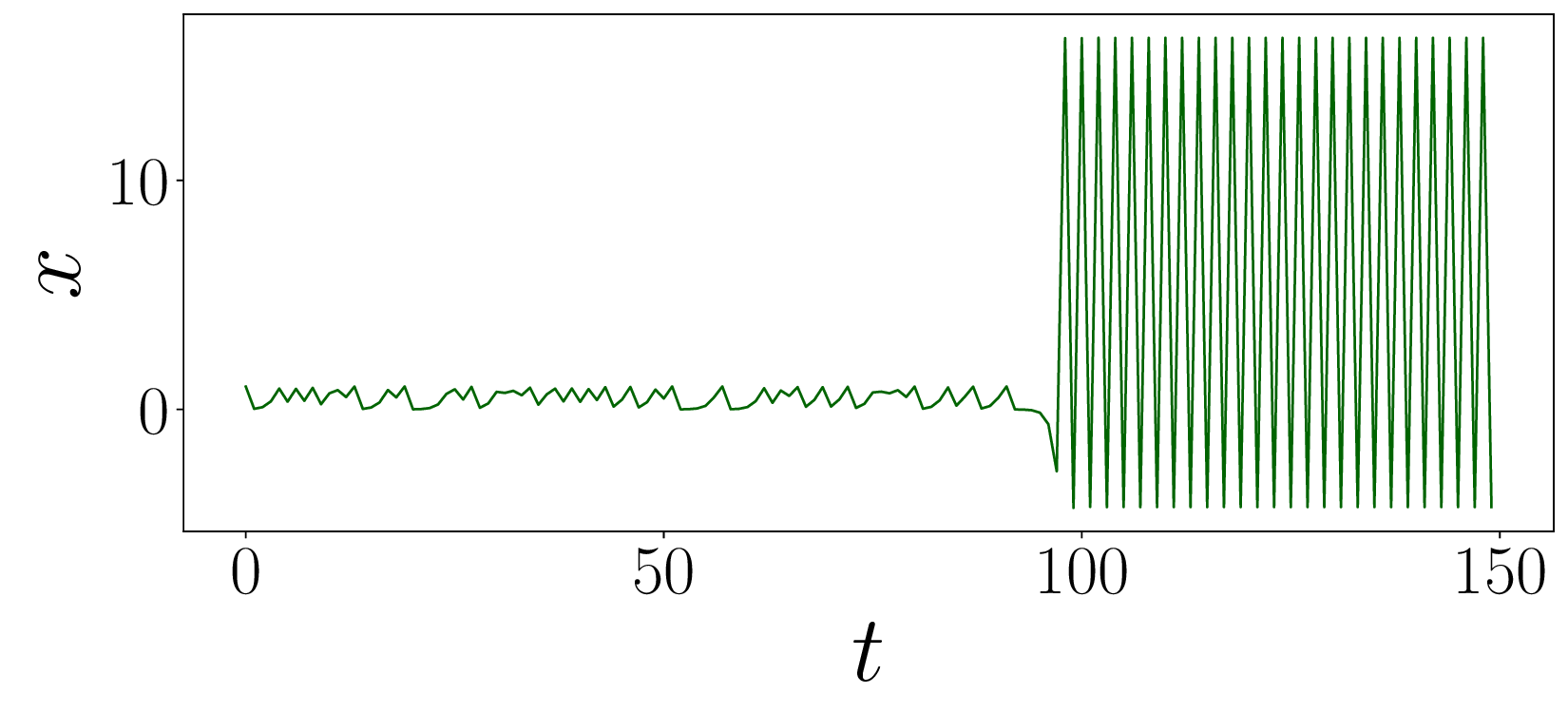}
    \caption{Time series of the trained reservoir map for $\mu_c < \mu$ correctly shows chaotic transient.}
    \label{figs7}
\end{figure}

We examine the RC map under a modified training dataset by using parameter values that are more widely spaced. In particular, the RC was trained on logistic map time series at $\mu \in \{3.8,3.85,3.9,3.95\}$, while keeping all other hyperparameters unchanged. The time series of trained map correctly shows the transient behavior, however instead of converging to a fixed point attractor, it settles to a stable periodic orbit of period 2 as shown in Fig. \ref{figs7}, still keeping the bifurcation structure before the crisis point the same. 

\begin{figure}[h!]
    \centering
    \begin{minipage}{0.48\linewidth}
        \centering
        \includegraphics[scale=0.26]{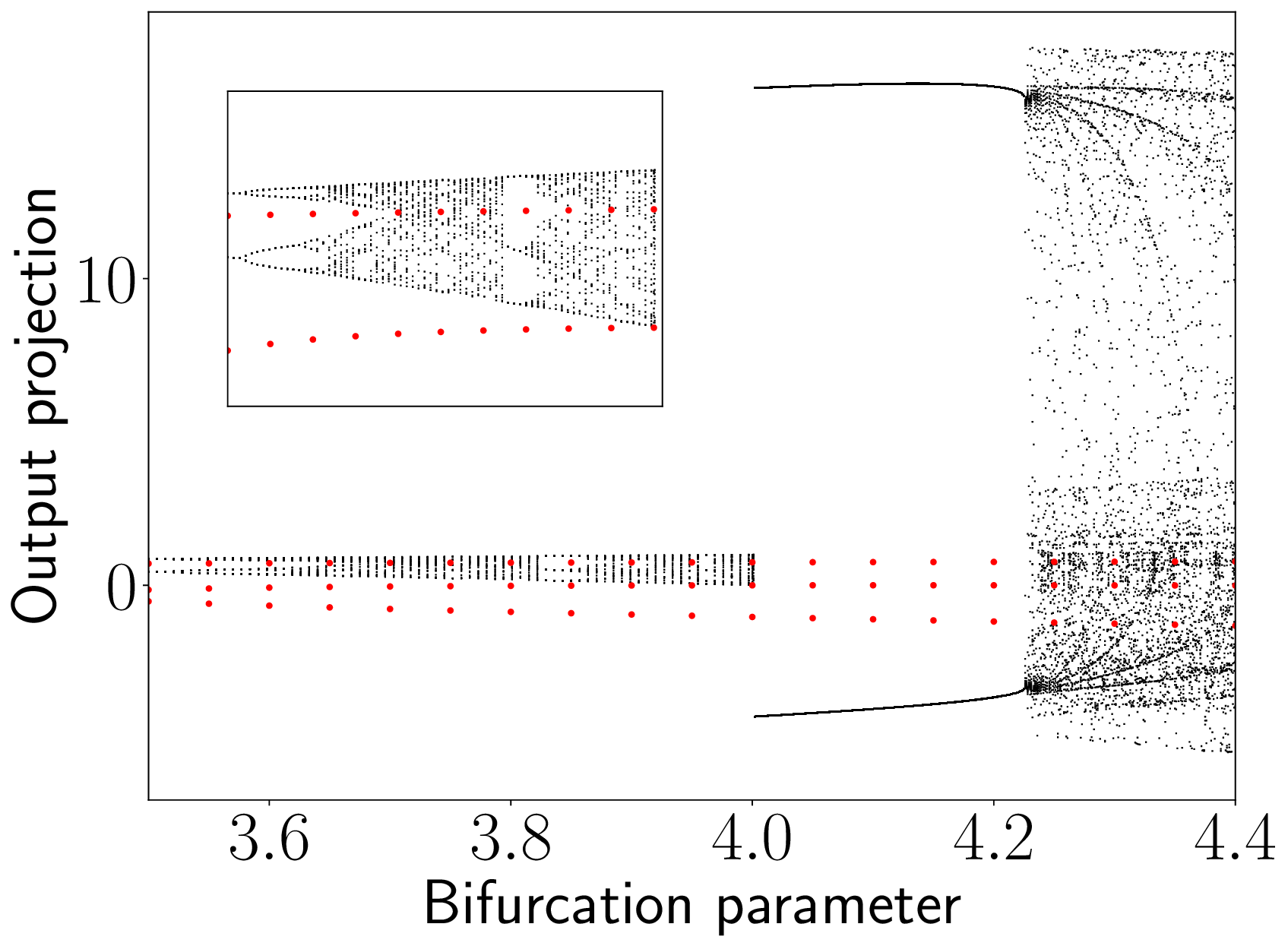}
        \caption{Collision of two unstable fixed points (red) with the chaotic attractor learned by the trained reservoir. Collision near $\mu = 3.7$ is responsible for attractor–merging crisis, while the collision near $\mu = 4$ causes boundary crisis. For $\mu > 4$, we observe a stable period-2 orbit.}
        \label{figs8}
    \end{minipage}
    \hfill
    \begin{minipage}{0.48\linewidth}
        \centering
        \includegraphics[scale=0.3]{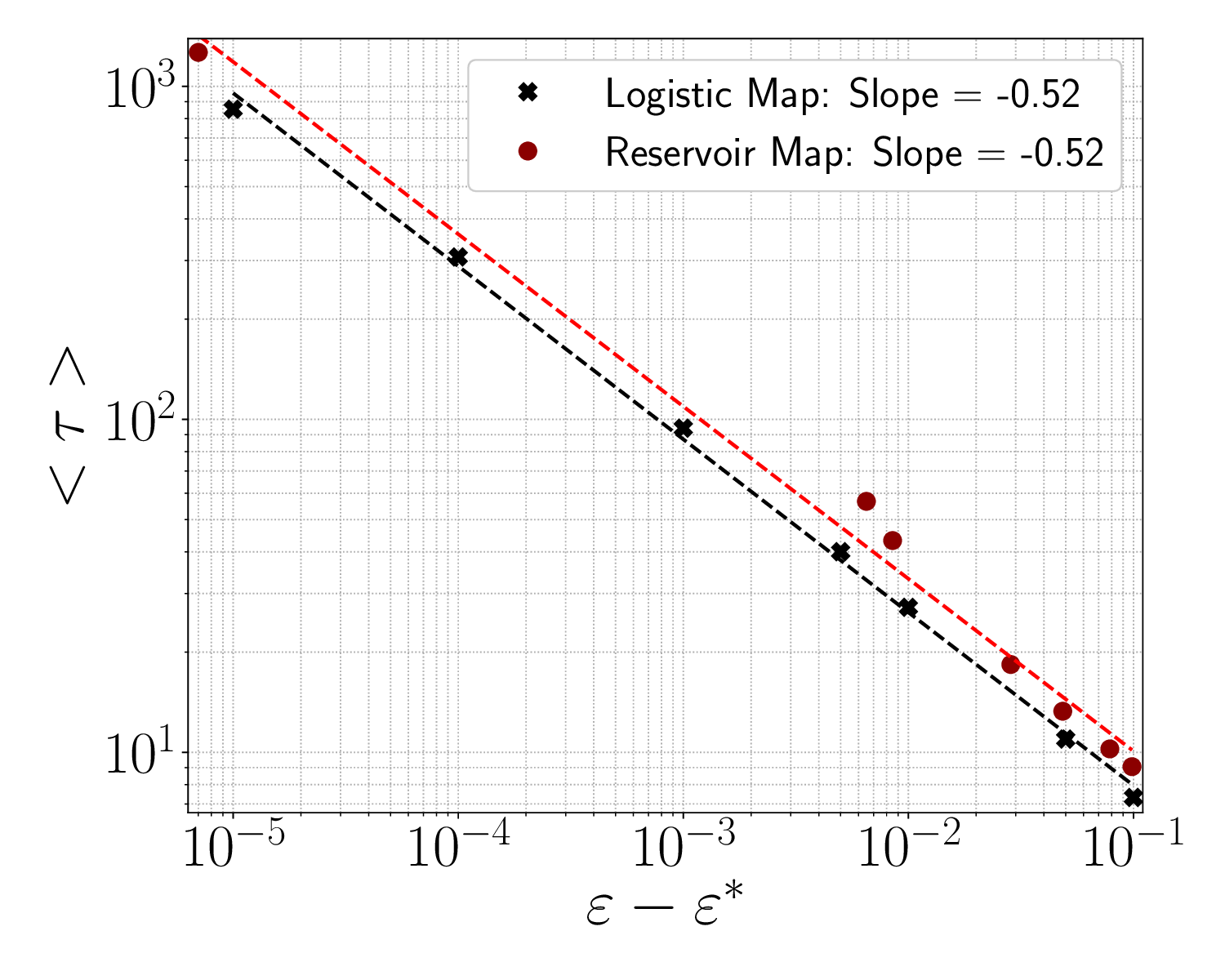}
        \caption{Log–log scaling plot: the trained reservoir map reproduces the scaling exponent of the logistic map. Averaging over $\tau$ uses $10^4$ different initial conditions.}
        \label{figs9}
    \end{minipage}
\end{figure}

Fixed point analysis shows the collision of the chaotic attractor with the unstable fixed points which depicts AMC and BC in the trained RC map (Fig. \ref{figs8}). The scaling exponents are also compared around $\mu = 4$ as depicted in Fig. \ref{figs9}. Because the training dataset lies farther from the crisis point, the trained RC map shows slight deviations from the logistic map; nevertheless, it remains a reliable representation for predicting the crisis.

\newpage 
\section{Training RC on H\'enon map}
We train the RC map on the two-dimensional H\'enon map \cite{PhysRevE.49.1094} in order to examine the challenges involved in understanding the mechanism of crises prediction in higher-dimensional systems. The H\'enon map is defined as
\begin{align*}
x_{n+1} &= a - x_n^2 + b y_n \\
y_{n+1} &= x_n.
\end{align*}
We fix $b = 0.3$ and treat $a$ as the bifurcation parameter. This map possesses two fixed points, both of which are saddles for $a > 0.3675$, as determined from Jacobian analysis. A boundary crisis occurs at $a_c \approx 1.426$: for $a < a_c$, a chaotic attractor exists, while for $a > a_c$ the attractor is destroyed and trajectories diverge after a finite transient.

\begin{figure}[h!]
    \centering
    \begin{minipage}[b]{0.48\textwidth}
        \centering
        \includegraphics[scale=0.24]{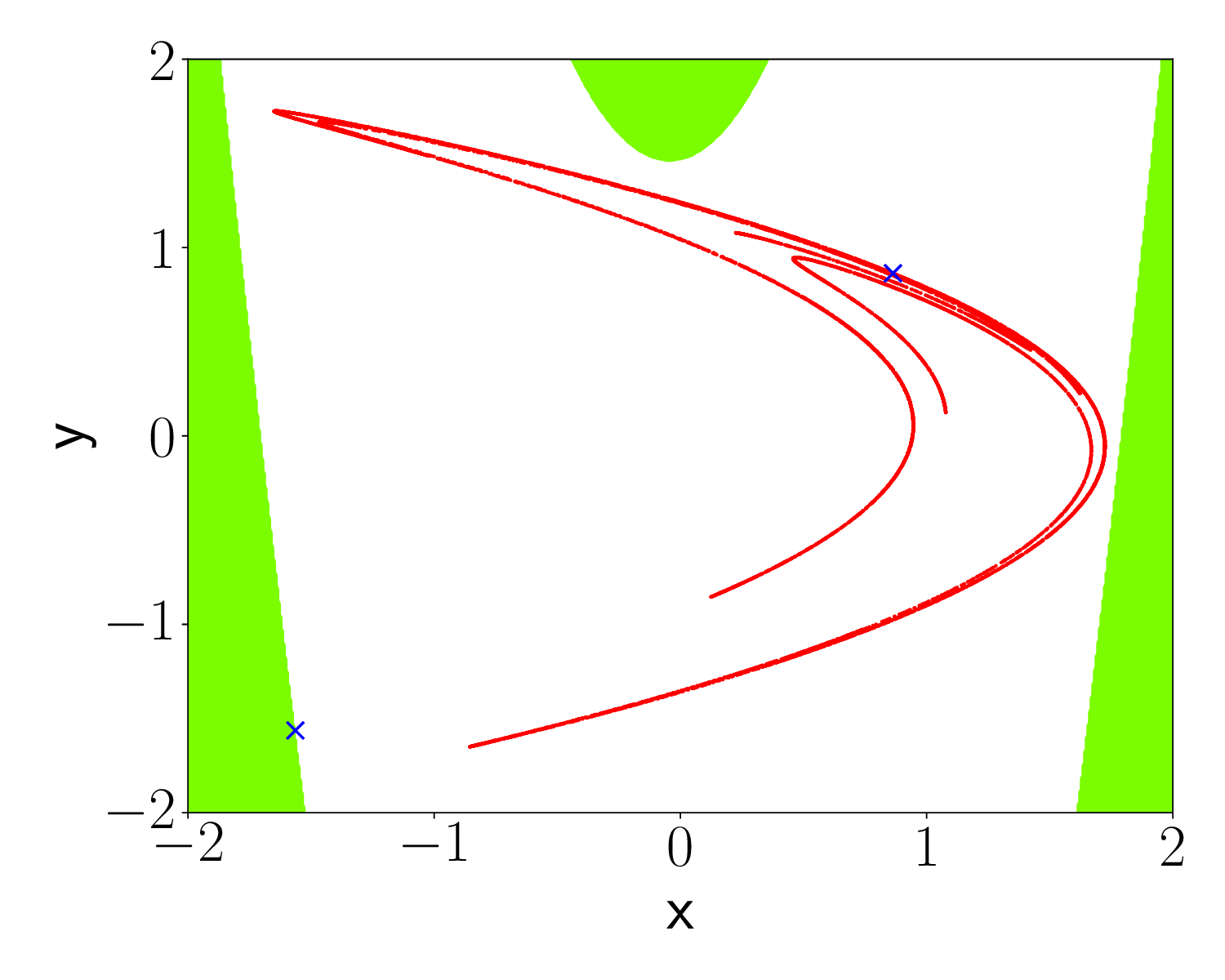}
        \caption{Chaotic attractor before the crisis ($a = 1.39$). Crosses mark the two saddle fixed points. The white region shows the basin of attraction of the chaotic attractor, while initial conditions in the green region diverge to infinity. The red trajectory depicts the chaotic attractor, which also corresponds to the unstable manifold of the saddle on the attractor.}
        \label{figs10a}
    \end{minipage}
    \hfill
    \begin{minipage}[b]{0.48\textwidth}
        \centering
        \includegraphics[scale=0.24]{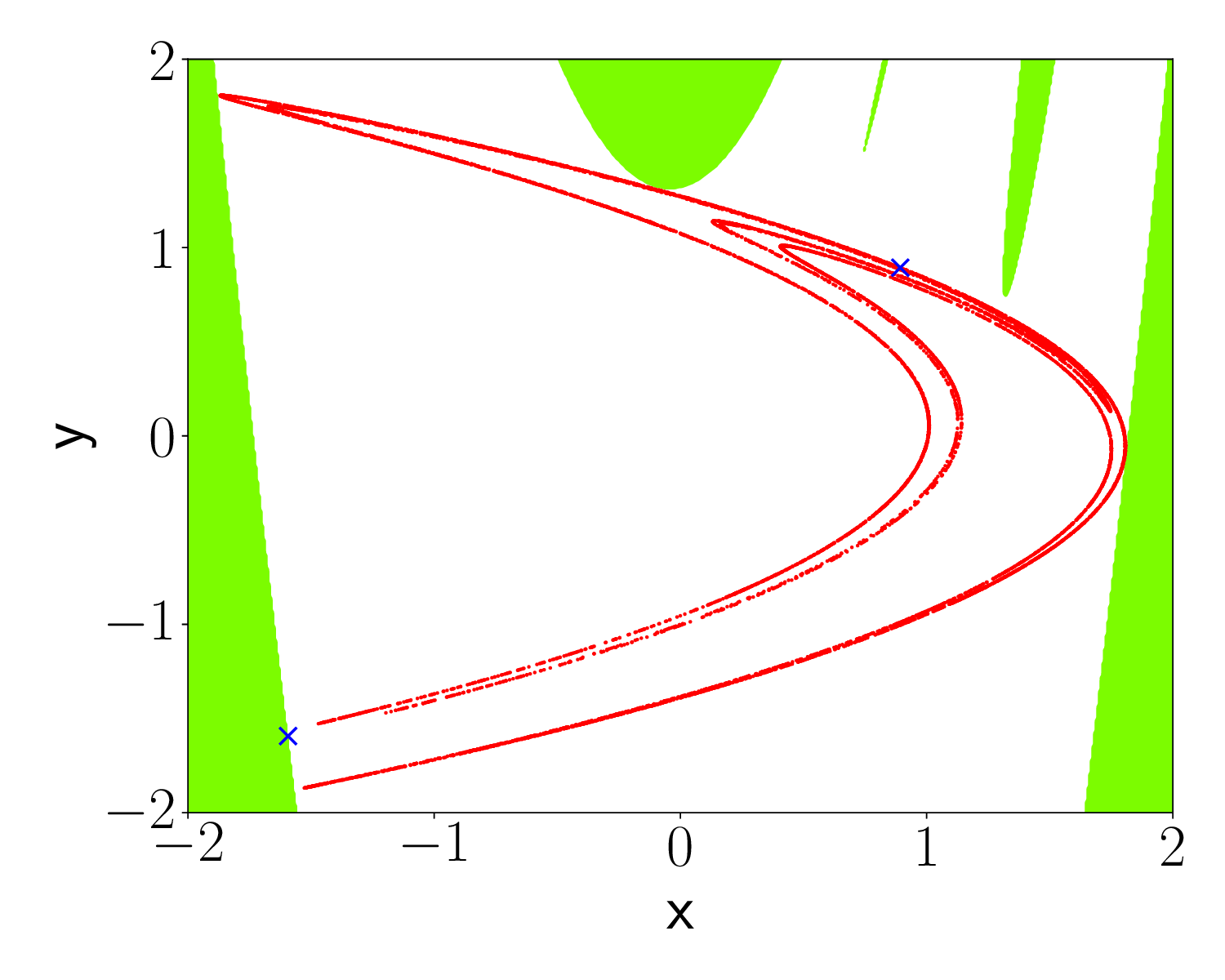}
        \caption{At the boundary crisis ($a = 1.424$), the chaotic attractor in red (unstable manifold of the saddle on the attractor) collides with the green basin boundary (stable manifold of the saddle on the boundary).}
        \label{figs10b}
    \end{minipage}

    \label{figs10}
\end{figure}

Mechanistically, a boundary crisis arises when the stable manifold of a saddle (which forms the basin boundary) collides with the unstable manifold of another (or the same) saddle that encloses the chaotic attractor. In one-dimensional maps, this reduces to the unstable fixed point colliding directly with the attractor. In higher-dimensional systems such as the H\'enon map, however, the situation is more intricate: the chaotic attractor is enclosed by the unstable manifold of a saddle, and the crisis occurs when this manifold collides with the basin boundary of the attractor, as illustrated in Fig. (\ref{figs10a}, \ref{figs10b}).

The RC map is successfully trained on H\'enon map time series at $a \in {1.38, 1.39, 1.4, 1.41}$ to reproduce the crises using the following set of hyperparameters $m = 400, \alpha = 0.86, b = 2.13, c = 1.15, \sigma = 0.15, \rho = 0.8, k_b = 1, \varepsilon_b = 0.3$. But unlike the case of one-dimensional systems, the collision between the fixed point and the chaotic attractor no longer happens. Since the trained RC map is 400-dimensional in our case, it becomes practically impossible to plot the stable and unstable manifolds to demonstrate the crises mechanism.

The situation is even more complicated in higher-dimensional systems such as the three-dimensional food chain model \cite{mccann1994}. There, boundary crisis involve the stable manifold of a saddle-type periodic orbit colliding with the chaotic attractor. In the RC map representation, the challenge is therefore twofold: identifying a saddle-type periodic orbit of unknown period, and then tracing its manifolds to verify their intersection at the crisis.

\nocite{*}


\begin{thebibliography}{99}

\bibitem{Tsironis2025}
G. Tsironis, ``Complex Systems and Machine Learning,'' in \textit{Artificial Intelligence and Complex Dynamical Systems}, Understanding Complex Systems, Springer, Cham (2025). \url{https://doi.org/10.1007/978-3-031-81946-9_1}.


\bibitem{Brunton2016}
S.~L. Brunton, J.~L. Proctor, and J.~N. Kutz, 
``Discovering governing equations from data by sparse identification of nonlinear dynamical systems,'' 
\textit{Proceedings of the National Academy of Sciences of the United States of America}, vol.~113, no.~15, pp.~3932--3937, 2016. 
doi: \href{https://doi.org/10.1073/pnas.1517384113}{10.1073/pnas.1517384113}.

\bibitem{Bury2023}
T. M. Bury, D. Dylewsky, C. T. Bauch, \textit{et al.}, 
``Predicting discrete-time bifurcations with deep learning,'' 
\textit{Nature Communications}, vol.~14, p.~6331, 2023. 
doi: \href{https://doi.org/10.1038/s41467-023-42020-z}{10.1038/s41467-023-42020-z}.

\bibitem{Choudhary2020}
A. Choudhary, J. F. Lindner, E. G. Holliday, S. T. Miller, S. Sinha, and W. L. Ditto, 
``Physics-enhanced neural networks learn order and chaos,'' 
\textit{Physical Review E}, vol.~101, no.~6, p.~062207, 2020. 
doi: \href{https://doi.org/10.1103/PhysRevE.101.062207}{10.1103/PhysRevE.101.062207}.

\bibitem{10.1063/1.5120538}
Z.-K. Gao, Y.-L. Li, Y.-X. Yang, and C. Ma,
A recurrence network-based convolutional neural network for fatigue driving detection from EEG,
\textit{Chaos} \textbf{29}, 113126 (2019).
\url{https://doi.org/10.1063/1.5120538}

\bibitem{LUKOSEVICIUS2009127}
M. Lukoševičius and H. Jaeger,
Reservoir computing approaches to recurrent neural network training,
\textit{Computer Science Review} \textbf{3}, 127-149 (2009).
\url{https://www.sciencedirect.com/science/article/pii/S1574013709000173}

\bibitem{doi:10.1126/science.1091277}
H. Jaeger and H. Haas,
Harnessing Nonlinearity: Predicting Chaotic Systems and Saving Energy in Wireless Communication,
\textit{Science} \textbf{304}, 78-80 (2004).
\url{https://www.science.org/doi/abs/10.1126/science.1091277}

\bibitem{10.1063/1.5039508}
Z. Lu, B. R. Hunt, and E. Ott,
Attractor reconstruction by machine learning,
\textit{Chaos} \textbf{28}, 061104 (2018).
\url{https://doi.org/10.1063/1.5039508}



\bibitem{PhysRevResearch.3.013090}
L.-W. Kong, H.-W. Fan, C. Grebogi, and Y.-C. Lai,
Machine learning prediction of critical transition and system collapse,
\textit{Phys. Rev. Res.} \textbf{3}, 013090 (2021).
\url{https://link.aps.org/doi/10.1103/PhysRevResearch.3.013090}

\bibitem{PhysRevE.110.034211}
D. Sisodia and S. Jalan,
``Dynamical analysis of a parameter-aware reservoir computer,''
\textit{Phys. Rev. E}, vol. 110, no. 3, p. 034211, Sep. 2024.
doi: \href{https://doi.org/10.1103/PhysRevE.110.034211}{10.1103/PhysRevE.110.034211}.

\bibitem{Hart2020}
A.~Hart, J.~Hook, and J.~Dawes,
``Embedding and approximation theorems for echo state networks,''
\textit{Neural Networks}, vol.~128, pp.~234--247, 2020.  
doi: \href{https://doi.org/10.1016/j.neunet.2020.05.013}{10.1016/j.neunet.2020.05.013}.

\bibitem{Bollt2021}
E.~Bollt, 
``On explaining the surprising success of reservoir computing forecaster of chaos? The universal machine learning dynamical system with contrast to VAR and DMD,''
\textit{Chaos}, vol.~31, no.~1, p.~013108, 2021.  
doi: \href{https://doi.org/10.1063/5.0024890}{10.1063/5.0024890}.


\bibitem{PhysRevLett.48.1507}
C. Grebogi, E. Ott, and J. A. Yorke,
``Chaotic Attractors in Crisis,''
\textit{Phys. Rev. Lett.}, vol. 48, no. 22, pp. 1507--1510, May 1982.
doi: \href{https://doi.org/10.1103/PhysRevLett.48.1507}{10.1103/PhysRevLett.48.1507}.

\bibitem{PhysRevLett.125.228701}
A.~Keane, B.~Krauskopf, and T.~M.~Lenton,
``Signatures consistent with multifrequency tipping in the Atlantic Meridional Overturning Circulation,''
\textit{Phys. Rev. Lett.} \textbf{125}, 228701 (2020).
\url{https://link.aps.org/doi/10.1103/PhysRevLett.125.228701}


\bibitem{PhysRevResearch.6.043194}
S.~Panahi, L.-W.~Kong, M.~Moradi, Z.-M.~Zhai, B.~Glaz, M.~Haile, and Y.-C.~Lai,
``Machine learning prediction of tipping in complex dynamical systems,''
\textit{Phys. Rev. Res.} \textbf{6}, 043194 (2024).
\url{https://link.aps.org/doi/10.1103/PhysRevResearch.6.043194}


\bibitem{xiao2021predicting}
Xiao, R., Kong, L.-W., Sun, Z.-K., and Lai, Y.-C.,
``Predicting amplitude death with machine learning,''
\textit{Physical Review E}, vol.~104, no.~1, p.~014205, 2021.
\href{https://doi.org/10.1103/PhysRevE.104.014205}{doi:10.1103/PhysRevE.104.014205}.

\bibitem{8016349}
G. E. Katz and J. A. Reggia,  
``Using Directional Fibers to Locate Fixed Points of Recurrent Neural Networks,''  
\textit{IEEE Transactions on Neural Networks and Learning Systems}, vol. 29, no. 8, pp. 3636--3646, 2018.  
doi: \href{https://doi.org/10.1109/TNNLS.2017.2733544}{10.1109/TNNLS.2017.2733544}.

\bibitem{directional-fibers}
Katz, G. (2023). \textit{Directional Fibers} [Computer software]. GitHub repository.  
Available at: \url{https://github.com/garrettkatz/directional-fibers}




\bibitem{RevModPhys.53.643}
J.-P. Eckmann,  
``Roads to turbulence in dissipative dynamical systems,''  
\textit{Rev. Mod. Phys.}, vol. 53, no. 4, pp. 643--654, Oct. 1981.  
doi: \href{https://doi.org/10.1103/RevModPhys.53.643}{10.1103/RevModPhys.53.643}.

\bibitem{Grebogi1985}
Grebogi, C., Ott, E., \& Yorke, J. A. (1985). Super persistent chaotic transients. 
\textit{Ergodic Theory and Dynamical Systems}, 5(3), 341--372. 
\href{https://doi.org/10.1017/S014338570000300X}{doi:10.1017/S014338570000300X}.

\bibitem{ananthkrishnan2001crises}
Ananthkrishnan, N., and Sahai, T.,
``Crises,''
\textit{Resonance}, vol.~6, pp.~19--33, 2001.
\href{https://doi.org/10.1007/BF02837669}{doi:10.1007/BF02837669}.

\bibitem{Ott2002}
E. Ott, \textit{Chaotic transitions}, in \textit{Chaos in Dynamical Systems},  
(Cambridge University Press, Cambridge, 2002), pp. 304–344.


\bibitem{Kong2021}
Ling-Wei Kong, Huawei Fan, Celso Grebogi, and Ying-Cheng Lai, 
``Emergence of transient chaos and intermittency in machine learning,'' 
\textit{Journal of Physics: Complexity}, vol. 2, no. 3, p. 035014, 2021. 
DOI: \href{https://doi.org/10.1088/2632-072X/ac0b00}{10.1088/2632-072X/ac0b00}.


\bibitem{supp}
See Supplemental Material at URL-will-be-inserted-by-publisher for more details.

\bibitem{mccann1994}
K.~Mccann and P.~Yodzis,
``Nonlinear Dynamics and Population Disappearances,''
\textit{Am. Nat.}, vol.~144, pp.~873--879, Nov. 1994, doi: 10.1086/285714.

\bibitem{Oliveira2022}
J.~A.~d.~Oliveira, H.~M.~J.~d.~Mendonça, V.~A.~Favarim \textit{et al.},
Eur. Phys. J. Spec. Top. \textbf{231}, 381 (2022).
doi:10.1140/epjs/s11734-021-00402-8

\bibitem{PhysRevE.49.1094}
Y.-C. Lai and C. Grebogi,
``Converting transient chaos into sustained chaos by feedback control,''
Phys. Rev. E \textbf{49}, 1094--1098 (1994).

\end{thebibliography}

\end{document}